%% file: backoff-measure.tex
\documentclass[5p,twocolumn]{elsarticle}

\usepackage{comment}
\usepackage{url}
\usepackage{color}
\usepackage{verbatim}
\usepackage{graphicx}
\usepackage[centerlast,it,IT,small]{subfigure}
\usepackage{morefloats}
\usepackage{float}
\usepackage{amsmath}
\usepackage{amssymb}
\usepackage{slashbox}
\usepackage{xspace}
\usepackage{tikz}
\usepackage{rawfonts}
\usepackage{color, colortbl}
\usepackage{times}
\usepackage[centerlast,font=small,format=plain,labelfont=bf,up,textfont=it,up]{caption}
\usepackage{array}
\usepackage{balance}

\DeclareCaptionType{copyrightbox}

\definecolor{Gray}{gray}{0.9}
\definecolor{mGray}{gray}{0.8}
\definecolor{dGray}{gray}{0.7}

\long\def\symbolfootnote[#1]#2{\begingroup%
\def\thefootnote{\fnsymbol{footnote}}\footnote[#1]{#2}\endgroup} 

\newcommand\eat[1]{}
\newcommand{\bi}{\begin{itemize}}
\newcommand{\ei}{\end{itemize}}
\newcommand{\im}{\item}
\newcommand{\eg}{{\it e.g.}\xspace}
\newcommand{\ie}{{\it i.e.}\xspace}
\newcommand{\etal}{{\it et.al.}\xspace}

\begin{document}

\begin{frontmatter}

\title{How Penalty Leads to Improvement: \\ a Measurement Study of Wireless Backoff}



\author[label1,label2]{Dmitriy~Kuptsov}
\ead{dmitriy.kuptsov@hiit.fi}

\author[label1,label2]{Boris~Nechaev}
\ead{boris.nechaev@hiit.fi}

\author[label2]{Andrey~Lukyanenko}
\ead{laser.ru@gmail.com}

\author[label1,label3]{Andrei~Gurtov}
\ead{gurtov@hiit.fi}

\address[label1]{HIIT, Finland}
\address[label2]{Aalto University, Finland}
\address[label3]{Center for Wireless Communication, University of Oulu, Finland}

\begin{abstract}

Despite much theoretical work, different modifications of backoff protocols in 802.11 networks 
lack empirical evidence demonstrating their real-life performance. To fill the gap we have set 
out to experiment with performance of exponential backoff by varying its backoff factor. Despite 
the satisfactory results for throughput, we have witnessed poor fairness manifesting in severe 
capture effect. The design of standard backoff protocol allows already successful nodes to remain 
successful, giving little chance to those nodes that failed to capture the channel in the beginning. 
With this at hand, we ask a conceptual question: {\em Can one improve the performance of wireless 
backoff by introducing a mechanism of self-penalty, when overly successful nodes are penalized 
with big contention windows?} Our real-life measurements using commodity hardware demonstrate 
that in many settings such mechanism not only allows to achieve better throughput, but also 
assures nearly perfect fairness. We further corroborate these results with simulations and
an analytical model. Finally, we present a backoff factor selection protocol which can be 
implemented in access points to enable deployment of the penalty backoff protocol to consumer
devices.

\end{abstract}



\end{frontmatter}



\input{intro.tex}
\input{algo.tex}
\input{impl.tex}

\input{result.tex}
\input{protocol.tex}
\input{discussion.tex}
\input{related.tex}

\input{conclusion.tex}

\bibliographystyle{abbrv}
{
\small \balance \bibliography{backoff-measure} 
}
\end{document}

%% file: intro.tex
\section{Introduction}


There are several well-known problems associated with the communication in wireless LANs (WLANs). In particular, since nodes in such networks use shared medium in an unlicensed radio spectrum for transmission of frames, collisions are possible. Measurement study of large scale enterprise WLANs~\cite{Cheng:2006:jigsaw} showed that nearly $15\%$ of sender-receiver pairs experience significant loss due to collisions. Another problem is fairness. The study in~\cite{duda:80211} showed that 802.11 networks have good short-term fairness when the number of contending stations is small, \eg 2, and becomes worse for an increasing number of stations. 

In this paper we tackle these two problems by proposing two novel backoff mechanisms for 802.11 networks. The protocols, although different in design, have a similar inspiration and a common goal---to increase throughput and improve fairness. The underlying principle of both schemes is to penalize overly successful nodes with large contention windows and accordingly reward unsuccessful nodes with small contention windows. We discover that unlike the greedy backoff scheme of standard 802.11 protocol, our protocols assure better fairness and higher throughput.

To evaluate effectiveness of the proposed designs we, unlike many other researchers~\cite{duda:sigcomm:2005, kwak:performance,vu:collision:model}, take a measurement approach. Thus, to conduct the study we implement two existing and two novel backoff protocols on commodity hardware. We consider the details of implementation, data collection and calibration techniques to be our second principle contribution. 

Next, we conduct a series of real-life experiments in various scenarios. The goal here is to explore how throughput and fairness of different protocols change when we vary such parameters as backoff factor, number of clients and lossiness of the channel. Our third principal contribution is in revealing performance trends and optimal configurations in real-life experiments, simulations and an analytic model.

To operate properly, the proposed algorithms require an accurate estimation of the number of active wireless stations attached to the access point. Our last contribution in this work is design and analysis of two metrics for estimating the number of active clients. We implement these metrics in a Linux-based wireless access point and evaluate the system performance in an office-like environment. 

To preview our results, we have found that in many settings the penalty mechanisms improve fairness greatly. At the same time, such improvement is also accompanied by considerable increase in throughput. To be more specific, our experimental work indicated the following:

\bi
\im In all the scenarios that did not include hidden terminals, we have witnessed that the backoff with penalty improves throughput by $100\%$ when compared to the standard 802.11 backoff. 
\im In many settings the penalty mechanisms allowed to achieve nearly perfect fairness. Even with RTS/CTS mechanism disabled, two hidden terminals using backoff with penalty achieved nearly perfect fairness under certain conditions.
\im The penalty mechanisms reduced collision rate on average from $0.3$ to $0.14$ in close proximity environment and from $0.29$ to $0.09$ in sparse deployment.
\im We have observed that the penalty mechanisms do not significantly increase the delays for UDP traffic, and preform much like standard backoff protocol.
\ei

The rest of the paper is organized as follows. In section Section~\ref{sec:protocols:desc} we review two existing and introduce two novel backoff protocols that will be studied in this paper. Section~\ref{sec:impl} discusses the implementation details, experimental environment, data collection and calibration process. We devote Section~\ref{sec:empirical} to our experimental findings and Section~\ref{sec:simulations} to corroborating these findings with simulations. In Section~\ref{sec:analytical} we outline the theoretical analysis. Section~\ref{sect:protocol} describes the protocol for estimating the number of active clients. In Sections~\ref{sect:discussion} and Section~\ref{sec:related} we discuss the implications of our findings and their correspondence with related work. We make concluding remarks in Section~\ref{sec:conclusion}. 

%% file: algo.tex
\section{Backoff protocols}
\label{sec:protocols:desc}

In this section we review the modifications of the backoff protocols we have experimented with. 

\begin{figure}[ht]
\includegraphics[width=3in, height=1in]{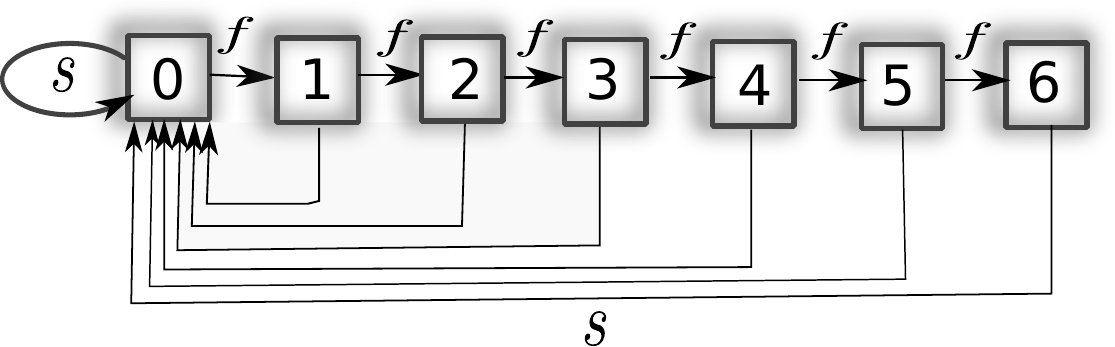}
\caption{Standard backoff}
\label{fig:original:backoff}
\end{figure}

\textit{Standard backoff with varying backoff factors.} This is the protocol that is currently in use in all 802.11 standards. We use this protocol as a benchmarking baseline and compare it with all other algorithms presented in this paper. The only modification of this protocol that we explore is varying backoff factors $r\in [1.2, 2.6]$. We discuss exact changes in Section~\ref{sec:impl}. Figure~\ref{fig:original:backoff} demonstrates how the standard backoff behaves upon succeeding ($s$) or failing ($f$) to send a packet. In case of failure, e.g. due to a collision, the protocol retransmits the packet and increases $i$ until it reaches $i=6$. If all $7$ retries fail, the packet is discarded and similarly to the result of a successful transmission, $i$ is set to $0$.

\begin{figure}[!hbt]
\includegraphics[width=3in, height=1.3in]{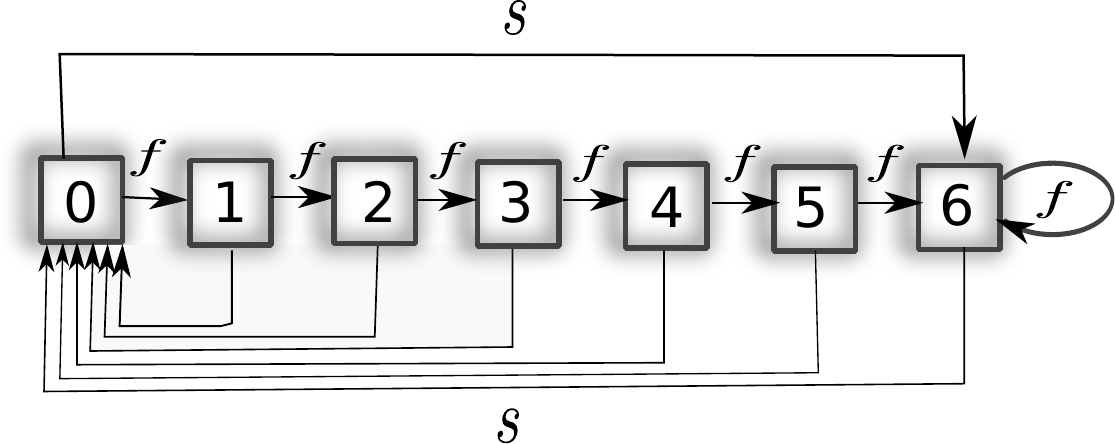}
\caption{Backoff with penalty}
\label{fig:reverse:backoff:1}
\end{figure}

\textit{Backoff with penalty}. This is the first novel protocol we propose in this paper. The idea behind it is to penalize the stations that always successfully transmit frames by throwing them to a state with a larger contention window, and at the same time allow the stations that are unsuccessful in transmissions to have smaller contention windows. As seen in Figure~\ref{fig:reverse:backoff:1}, this is achieved by putting the node which was in state $0$ and had a successful transmission into state $6$. Such a modification increases the chances of each station to seize its portion of wireless medium. Similarly to standard backoff, after 7 unsuccessful retries the mechanism retreats to state $0$. 

\begin{figure}[ht]
\includegraphics[width=3in, height=1in]{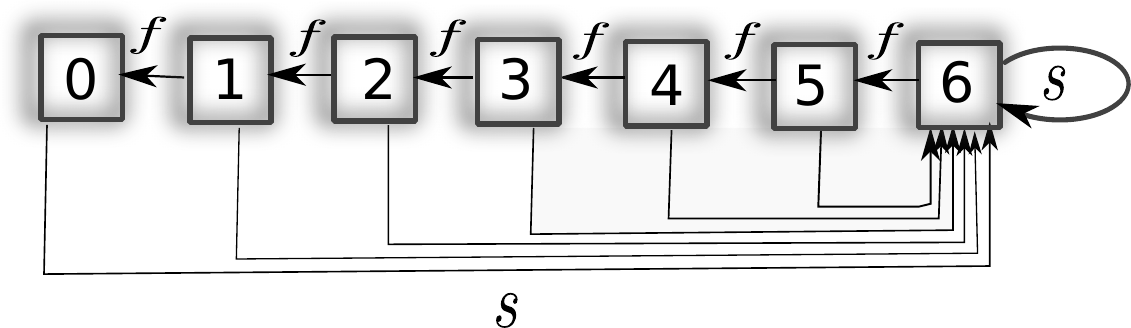}
\caption{Rollback backoff}
\label{fig:reverse:backoff:2}
\end{figure}

\textit{Rollback backoff}. In contrast to backoff with penalty, in our second proposed algorithm (rollback backoff) nodes always start with the state that has the largest contention window, and only nodes that are unsuccessful are rewarded with smaller contention windows. The state transition diagram for this protocol is presented in Figure~\ref{fig:reverse:backoff:2}. Unlike the previous two protocols, 7 failed retries set the state 	to $6$.

\textit{Backoff with fixed contention window}. This protocol is different from all protocols described above in that it has a single state. In other words, the nodes have up to $7$ retries (this is common for all protocols discussed in this paper) but after each failure or success the contention window remains unchanged. Though as we show later, optimal window size should depend on the number of active stations. Contention window sizes that we have used in our experimental work are identical to those presented in Table~\ref{tbl:ifs}.

%% file: impl.tex
\section{Implementation and data collection}
\label{sec:impl}

\begin{figure*}[!thb]
	\centering
        \subfigure[Close proximity setting: nodes are localized in one place]{
            \label{fig:testbed:simple}
            \includegraphics[width=3in, height=2in]{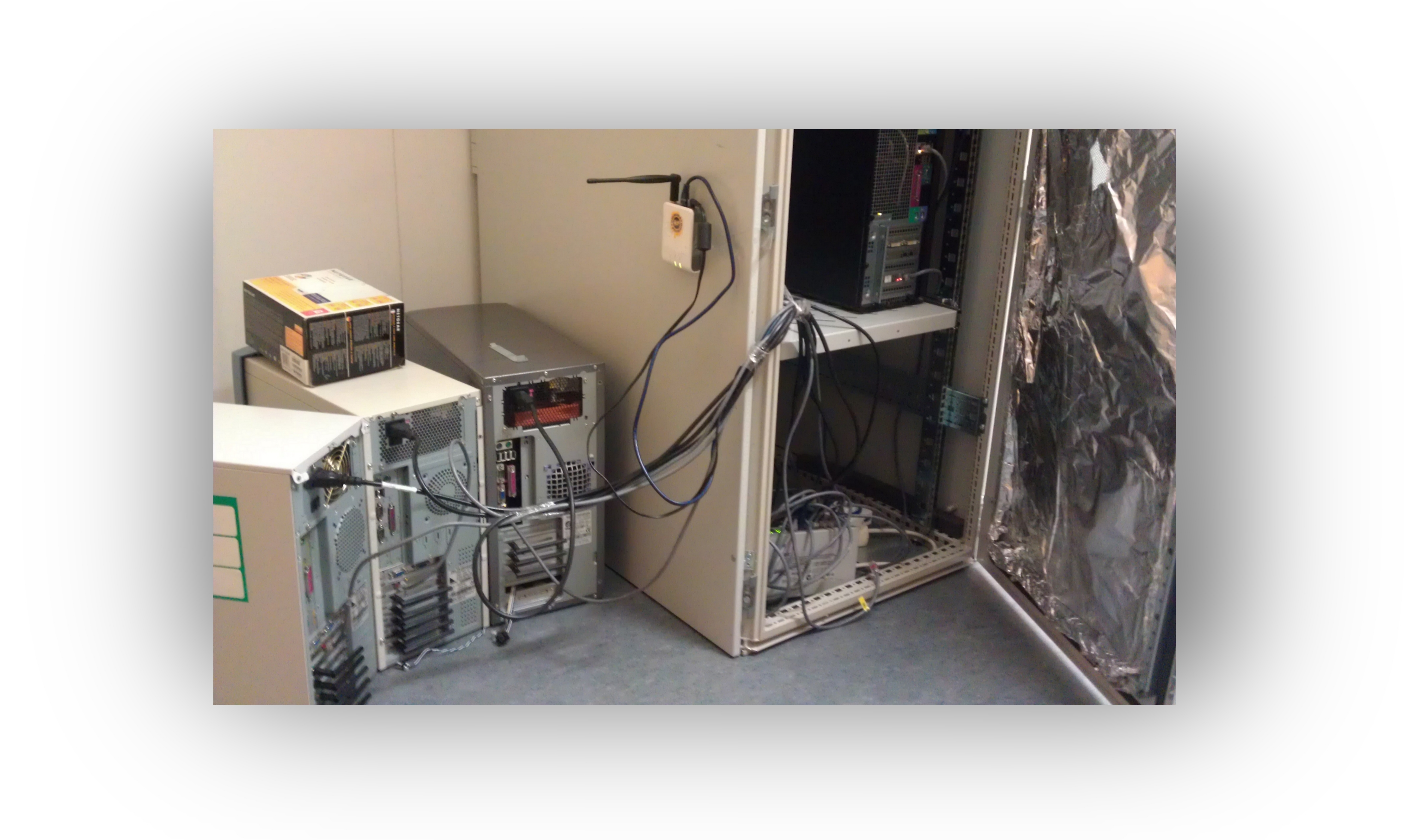}
        }
		\subfigure[Sparse deployment: nodes are scattered around the office]{
           \label{fig:office:map}
           \includegraphics[width=3.5in]{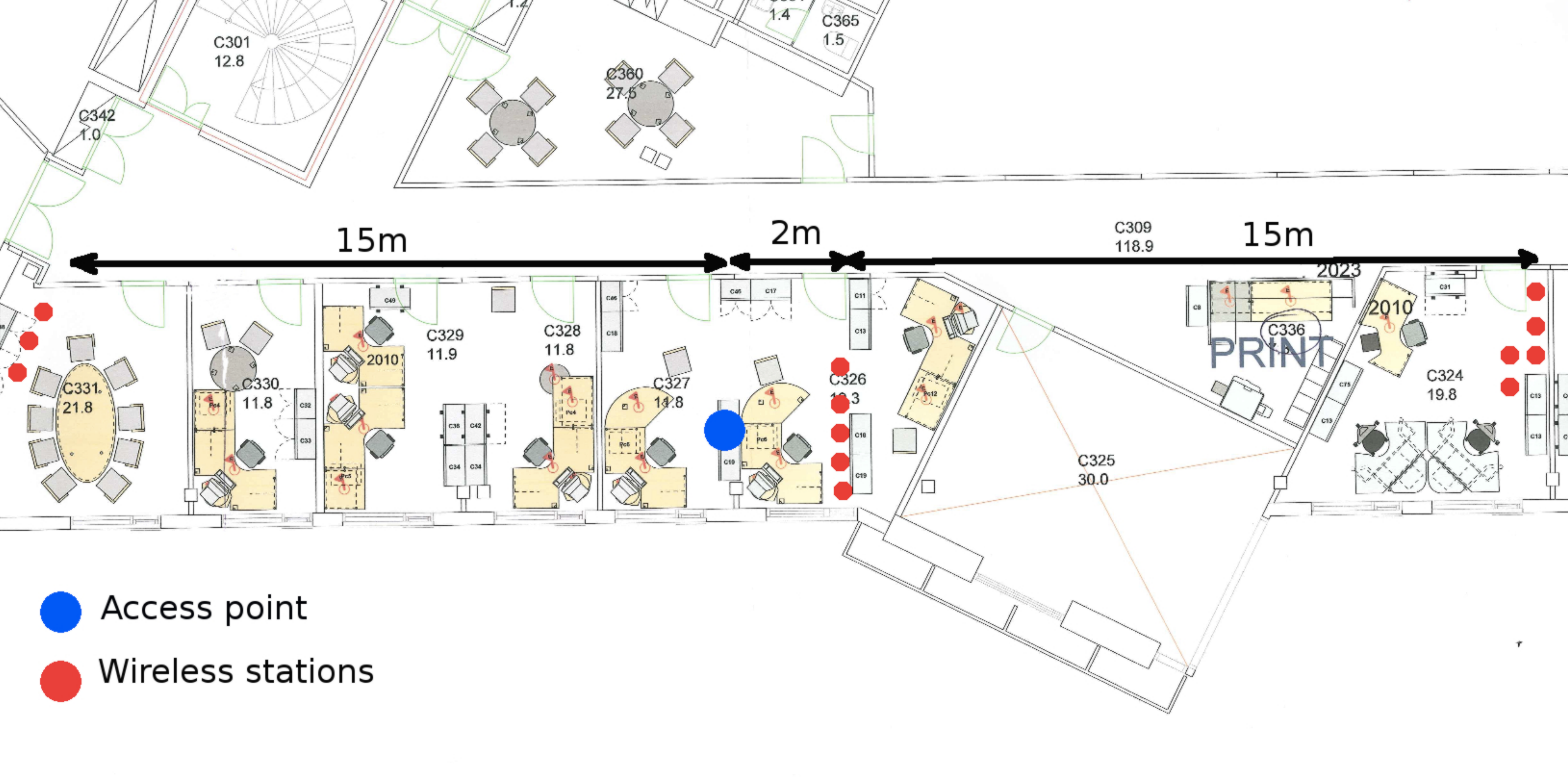}
        }
    \caption{
       Experimental testbeds
     }
    \label{fig:testbed:variants}
\end{figure*}

In this section we describe the Linux implementation of backoff protocol for 802.11 networks, our experimental testbed
and the way the data was collected and calibrated. 

\subsection{Implementation}

Most, if not all, hardware manufacturers implement such low level protocols as 802.11 backoff in firmware or in proprietary 
kernel modules. And typically, the firmware is shipped as a binary microcode that is loaded by OS drivers directly into device 
memory. This significantly complicates any changes to default protocol behavior. Fortunately, research community and open source 
enthusiasts provide a great opportunity to explore the internals of such software. 
In our implementation we have used open source firmware \cite{openfwwf} for Broadcom B43 wireless cards as a basis for experimental work. 
The microcode is written in assembly language and features implementation of standard 802.11g Medium Access Control (MAC) mechanisms for Broadcom/Airforce chipsets. 


Without going into various peculiarities we will instead describe what noticeable changes we have introduced to the firmware. The task was to modify the 
exponential backoff, as well as random backoff, and the way it is calculated in firmware. In original firmware the contention window is calculated as follows. Initial contention window $CW_{min}$, which is also the minimum one, is set to a default value of $31$, and the maximum is bounded with 
$CW_{max}=1023$. Whenever the collision is detected the client increases the window by a factor of $2$. 
The contention window is increased by shifting the value in the register \small\textit{CUR\_CONTENTION\_WIN}\normalsize~to 
the right. To ensure that the value of current contention window will not exceed the maximum 
allowed value $1023$, bitwise \textit{AND} operation with the \small\textit{MAX\_CONTENTION\_WIN}\normalsize~value is performed.
We introduced several changes to the above scheme.

\textbf{Non-standard backoff factors}. Instead of using a fixed backoff factor $r=2$, we allow  $r\in \mathbb{R}$ 
and experiment with the values in $[1.2, 1.3, \ldots 2.6]$. In our code $CW_{min}$ is reduced from $31$ to $15$, since 
IEEE 802.11g uses the latter. We also removed $CW_{max}=1023$, which was bounding the growth of $CW$. 
Thus, in our implementation contention window changes according to $CW_i=CW_{min}*r^i-1$, where $i \in [0, 6]$ is the retransmission counter
($i=0$ means that the packet is being sent for the first time). For the experiments with fixed contention window, we have set identical values to all $7$ states. For example, for the experiments with $3$ clients the contention window for all states was set to $21$~\cite{duda:sigcomm:2005}.

 Using floating point factors in original assembly code as above would require the support for floating point multiplication in hardware, which is 
of course missing. As a workaround we have done the following. First, using trial and error method we have found the offset in 
memory (note the available shared memory in the devices is just 4KB) which was not used by any part of the code. Second, for every $r$ 
we have precomputed possible contention window sizes and corresponding bit masks (we will discuss the need for such bit masks 
later). The mask was calculated as the minimum value $2^x-1$ which is larger than the corresponding contention window. 

Third, the client keeps track of backoff counter variable which is increased by $1$ whenever the collision occurs or resets it to $0$ whenever the contention window returns to the minimum $CW_{min}$. Observe, that the actual minimum contention window doesn't depend on $r$ and is always equal to 15. Also, for $r=2$ our scheme is identical to the one used in original IEEE 802.11g protocol.

Finally, we have changed the way random backoff value from the interval $[0,CW_i]$ is selected in the firmware. In the original version this 
was achieved using a bitwise \textit{AND} operation of the pseudo-random number generated in \small\textit{SPR\_TSF\_Random}\normalsize~register 
with the value of current contention window stored in \small\textit{CUR\_CONTENTION\_WIN}\normalsize. This allows to trim the 16-bit random number 
to the number of bits in the current contention window and thus obtain a random number in $[0,CW_{min}*2^i-1]$. Though since in our case the 
values of contention windows are not always powers of $2$, the above approach wouldn't work. 
In principle, it is possible to obtain the required random value in $[0, CW_i]$ by taking $(\small\textit{SPR\_TSF\_Random}\normalsize~mod~CW_i)$. Since 
division operation is absent in the microcode instruction set, we can subtract $CW_i$ from \small\textit{SPR\_TSF\_Random}\normalsize~until the result is
less than $CW_i$. We implemented this algorithm, but found that big number of frequent subtraction operations significantly degrades overall performance.
Thus, we switched to a more efficient solution. Given a $CW_i$ we do bitwise \textit{AND} operation between \small\textit{SPR\_TSF\_Random}\normalsize~and
the precomputed mask equal to the minimum value of $2^x-1$ bigger than $CW_i$. This gives us a random number $H \in [0, 2^x-1]$. If $H \le CW_i$, then 
we have found the sought random backoff value. Otherwise, if $H \in (CW_i, 2^x-1$], we repeat the above operations of taking a new 16-bit random number and
applying bitwise \textit{AND} with the mask, until $H \le CW_i$. Empirical evidence shows that the loop doesn't cause any noticeable performance issues.

\textbf{Non-standard state transition}. To implement rollback backoff and backoff with penalty we have also changed the way the clients select backoff states whenever collision happens. For instance, for the rollback backoff protocol instead of starting from state $0$ and increasing the backoff counter by one upon every collision, the clients start from state $6$ (maximum contention window) and decrement the counter on each retry. 

For backoff with penalty we have also added logical check to a callback which is invoked after successful frame transmission. If the packet was successfully transmitted at the first attempt then the state becomes $6$, otherwise the state is set to $0$.

\subsection{Experimental environment}

Our testbed comprised only commodity hardware. We have selected inexpensive (\$4 each) 802.11g wireless cards, 4 commodity computers, a 100/10 Base-T switch and single wireless access point running OpenWRT Linux, which also supported 802.11g standard. The variants of our testbed are shown in Figure~\ref{fig:testbed:variants}.

We have dedicated a single computer to play the role of a \textit{master node}. This node was responsible for sending commands to slave nodes to start experiments and also was participating in receiving and sending test traffic from and to slave nodes over wireless interfaces. The other 3 machines were used as \textit{slave nodes}. These nodes were provisioned with a single wired connection each and 5, 4 and 3 wireless cards correspondingly.

Since slave nodes had multiple interfaces, we configured these Linux boxes with policy based routing rules. This allowed us to send specific traffic through a specific interface. For instance, all control traffic such as commands and calibration packets (discussed below) was sent to wired interfaces, while wireless cards were performing experimental bulk transfers. Such setup allowed us to avoid interfering control traffic with experimental wireless flows. Since the experiments were conducted in an office-like environment, we have configured the wireless network with a channel that was least used and thus most probably not overlapping with other channels.

\begin{figure*}[!thb]
\begin{minipage}[b]{0.5\linewidth}
\centering
\includegraphics[width=1.8in]{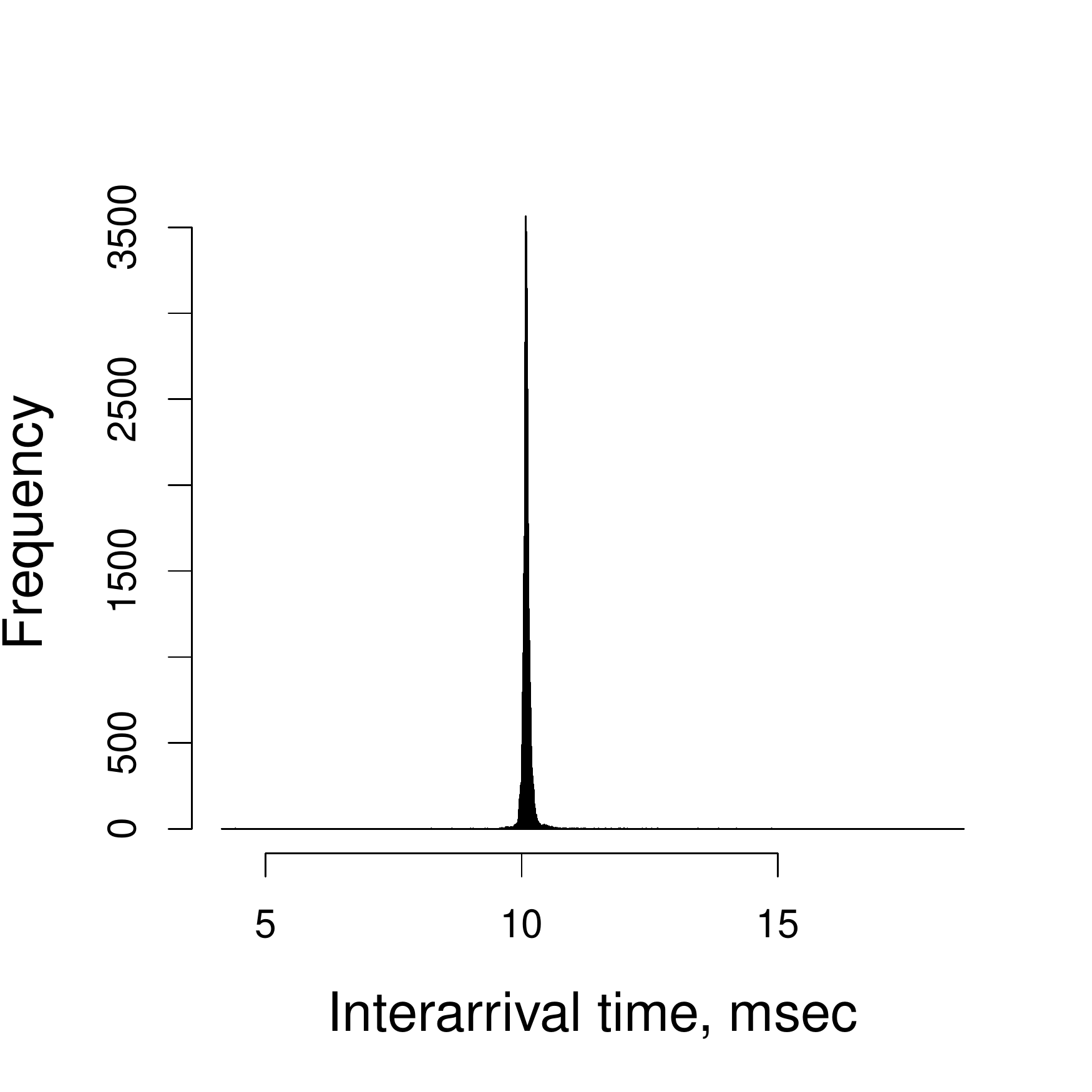}
\caption{Inter-arrival time between consecutive beacons.}
\label{fig:beacons-ivals}
\end{minipage}
\hspace{0.5cm}
\begin{minipage}[b]{0.5\linewidth}
\centering
\includegraphics[width=1.8in]{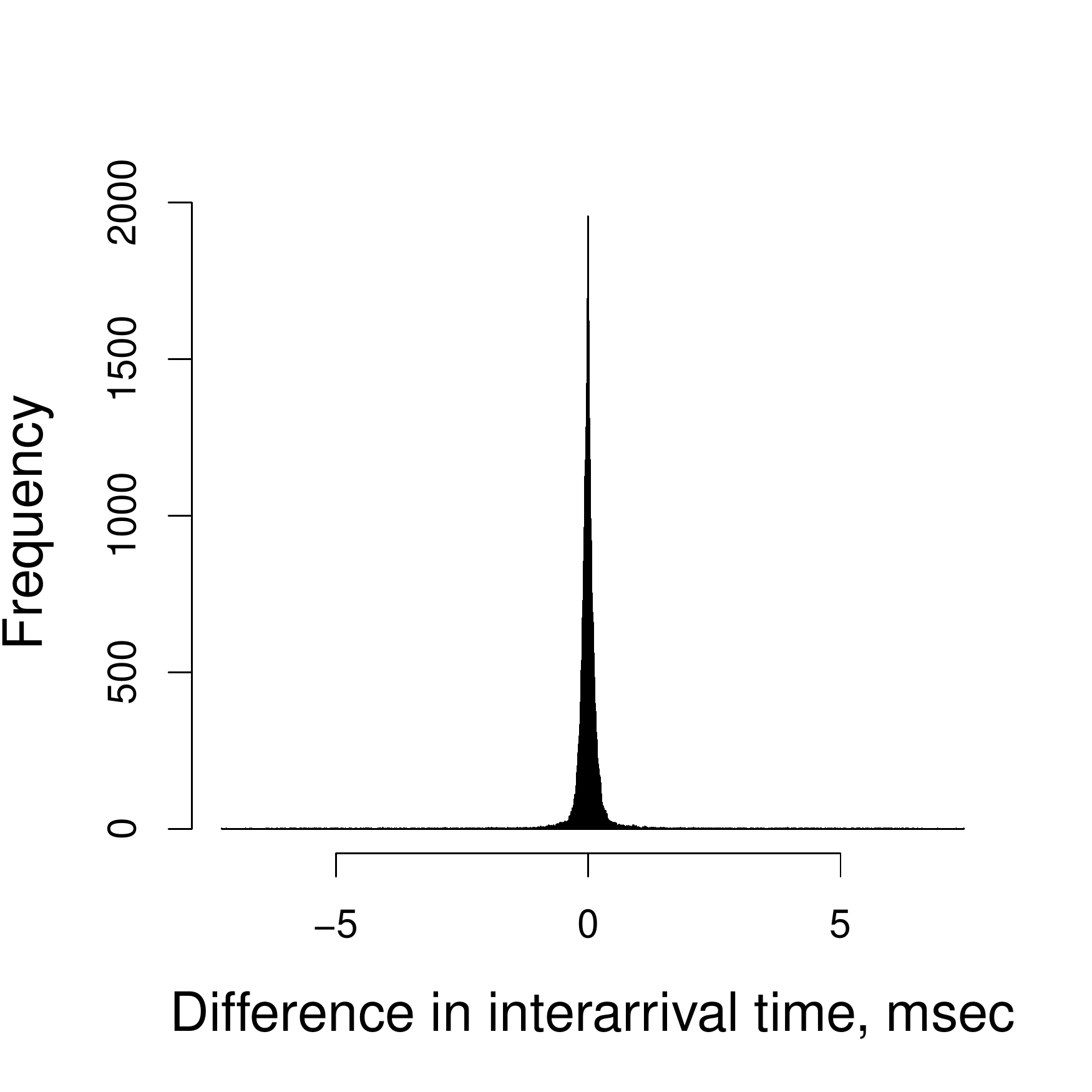}
\caption{Difference in beacons inter-arrival times for different machines.}
\label{fig:beacons-diffs}
\end{minipage}
\end{figure*}

\subsection{Data collection and calibration}

Our experiments consisted of three traffic types: bulky TCP upload and download streams, and delay sensitive UDP flows. 
We have generated the first traffic pattern with the transfers of 64 MB files using Linux \texttt{Wget} utility. Here the
master node was downloading the files from the slaves. Note that from the perspective of slave nodes (stations) this is the 
upload traffic. The second traffic pattern was generated using similar file sizes,
but now the master node was uploading the file to the slave nodes using \texttt{scp} utility (hence we call it download 
traffic in Section~\ref{sec:download}). We could implement a simple custom  
application running on top of UDP to get more control over rate limiting and congestion control dynamics. Though
we chose to use TCP, since it's the most popular transport protocol for performing large file transfers. 
We have used UDP in the experiments involving measurement of delays. In particular we have used UDP flows to emulate delay sensitive voice-over-IP sessions. We have implemented a custom application that was periodically 
(every $10 ms$) sending $100$ bytes packets from slave nodes to the master node. Using a socket option the master node 
was registering timestamps of each received UDP frame. This allowed us to achieve the required precision in measurements. 

We instructed the kernels on slave machines to log on per packet basis the information about number of retries, 
acknowledgment flags, packet size, used contention window and backoff interval. In doing so we have encountered a 
problem with Linux kernel which did not allow us to log this information too frequently. To overcome the 
issue we have recompiled the kernel with an increased ring buffer size for debug messages
and also increased kernel \textit{printk} rate limit. Upon receiving packet transmission status notification from firmware, 
the kernel registers the event and logs it into ring buffer. The ring buffer is periodically (every $0.1$ seconds)
read and dumped into a file. Each event was also flagged with the wireless interface ID and a timestamp. For the latter we found it easier to use 
time elapsed since machine boot up rather than time since the Unix epoch. 

Upon inspecting our logs, we discovered that some of them still contained few errors of two different types. First is 
transmission error reported by the wireless card driver. Unfortunately we failed to identify the cause for the error. 
Second is the kernel debugging error mentioned above---apparently even after tweaking kernel parameters, it still 
didn't manage to log all events and occasionally reported that some of them were suppressed. We estimated share of these
two errors relative to the number of benign log events and found it to be negligible, well below 1\%. Hence, we 
decided to ignore erroneous log messages by discarding them.

In total we used 12 wireless cards installed on three slave nodes. For majority of experiments we have used 3, 6, 9 and 12 
concurrently active clients. We have balanced the usage in such a way that for any number of active clients we have 
employed all three slave nodes in our testbed. Our eventual goal was to study combined 
performance of all active clients, which required merging logs recorded on different machines. Since clocks on
the machines were not in sync, we had to find a way to correctly align our logs. The solution was to send calibrating beacons 
from the master node to all slave machines via wired interfaces. In principle, it would have sufficed to send a single 
beacon in the beginning of each experiment, which slave nodes would have recorded as a reference time frame. Then subtracting
this value from each packet's timestamp would yield a relative packet's timestamp in the merged log file.
Alas, this solution is not perfect, since in prolonged experiments clock drift among different machines could 
cripple relative packet timings by putting some of them unduly further into the future or the past. To eliminate the
effect of clock drift we instructed the master node to send beacons periodically, with an interval of 10 msec. This made it
possible to do re-alignment on short timescales.

Another task where we utilized beacons is splitting logs into bins. Binning allowed us to do fine-grained analysis of 
various performance metrics and trace evolution of different relevant parameters over time. We found it convenient to
perform binning by tying bin size to a number of consecutive beacons. For instance, 10 beacons would correspond to 
a 100 msec bin, 100 beacons - to a 1 sec bin, etc. 

Even though the beacons were sent with a strict 10 msec interval, there is no guarantee that they were recorded by slave
nodes with exactly the same intervals. Perfect inter-arrival time between beacons could be skewed by various network, 
NIC or OS effects. Incorrect beacon inter-arrival times could result in imperfect binning and thus undermine any analysis
that relies on the assumption of constant bin size. To assess the possible skew in beacon timestamps, we calculated 
beacon inter-arrival times for all logs. A typical distribution is shown in Figure~\ref{fig:beacons-ivals}. To our relief,
inter-arrival times turned out to be sharply clustered around 10 msec. Though the figure still shows rare outliers. This
could lead to drift in cumulative beacon intervals among several machines. However, Figure~\ref{fig:beacons-diffs} eliminates
this concern. To make the plot we calculated differences between respective beacon inter-arrival times on different machines.
The facts that the distribution is centered at zero, highly clustered and symmetric proves that bins calculated based on
beacons are staying equally sized in the long run.

After devising start and end timestamps of each bin, we traverse our logs and using all packets falling into
it calculate number of successfully transmitted packets, number of failures and the total number of packets 
in the bin. The set of these three parameters is also calculated for each client participating in the experiment.

The last bit of calibration that we performed before moving to data analysis is truncating our logs to all-active periods.
Packet sending may have started at different times on different clients. Also, some clients may have finished sending
faster than the others. In our analysis we were interested in those periods when all clients were active, since this
would give us confidence that all of them were involved in competing for the channel. To identify such period, 
we calculated the maximum timestamp of the first full-size (1540 bytes) packet among all clients and the minimum 
timestamp of the last one. Then, using these two values we discarded all packets before and after
them respectively. We made sure that the remaining periods were big enough to provide meaningful data for our analysis.

%% file: result.tex
\section{Experimental results}
\label{sec:empirical}

\begin{figure}[t]
	\centering
        \subfigure[Backoff with penalty]{
            \label{fig:tput:close:enabled:gt}
            \includegraphics[width=1.5in]{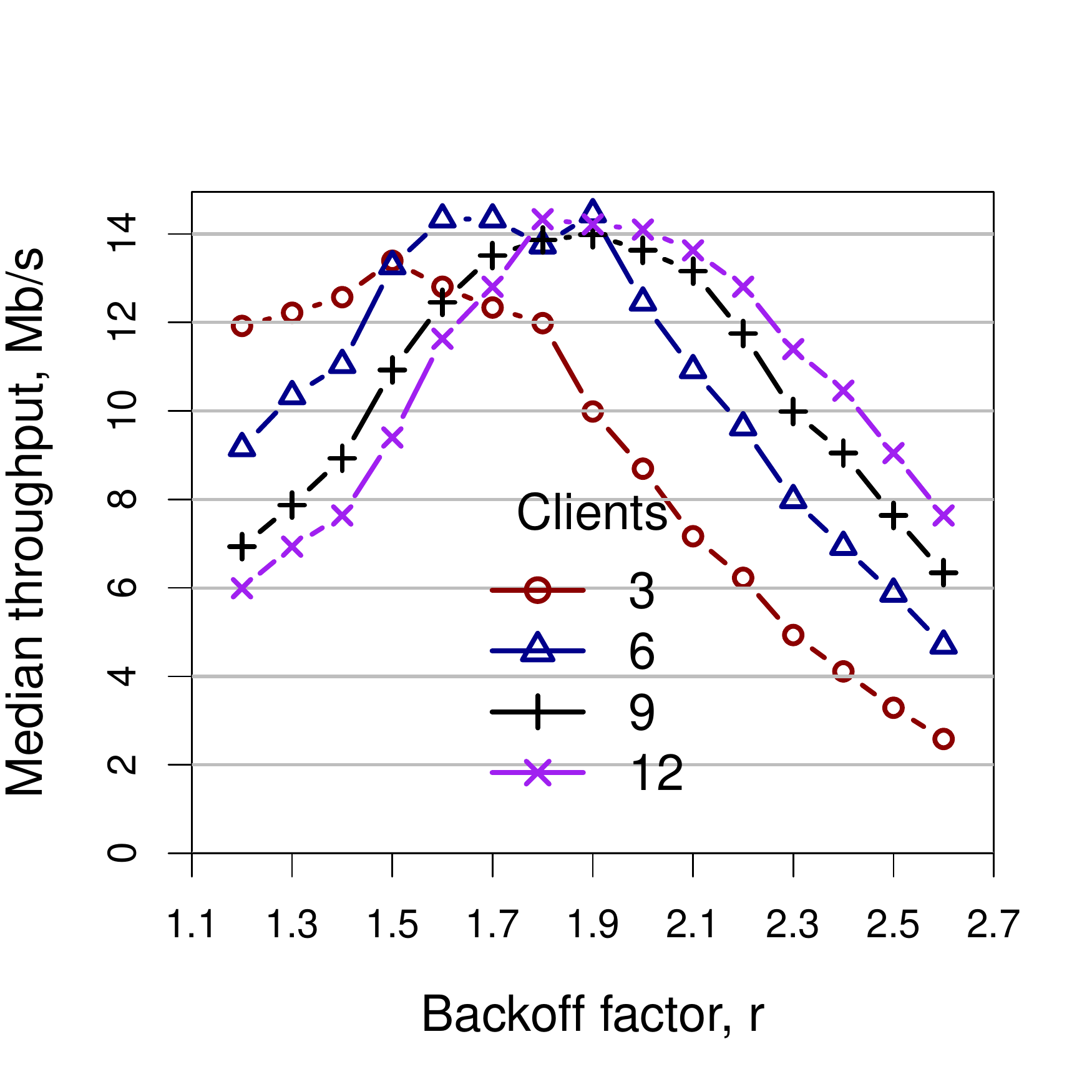}
        }
		\subfigure[Rollback backoff]{
           \label{fig:tput:close:enabled:roll}
           \includegraphics[width=1.5in]{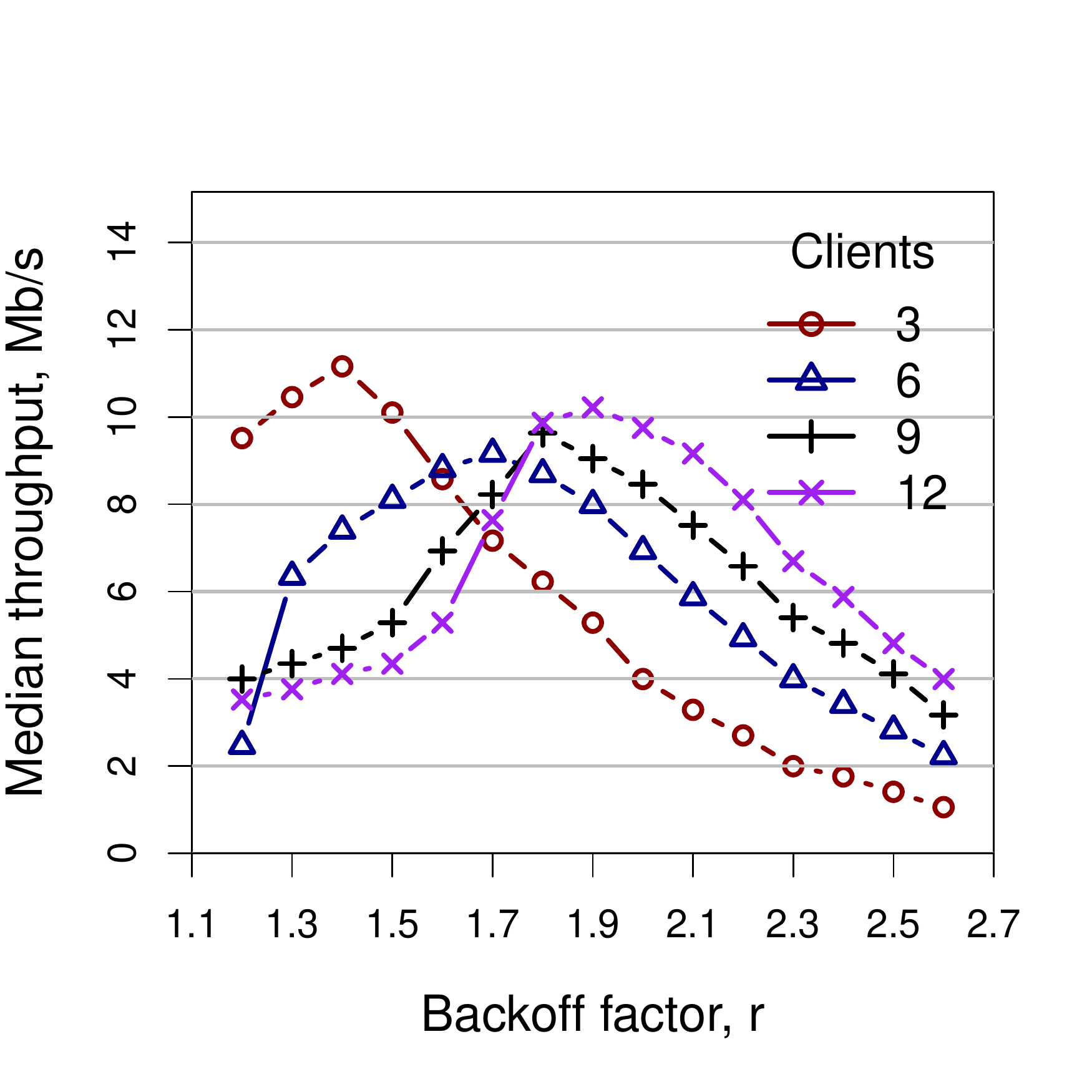}
        }
		\subfigure[Backoff with fixed contention window]{
            \label{fig:tput:close:enabled:fixed}
            \includegraphics[width=1.5in]{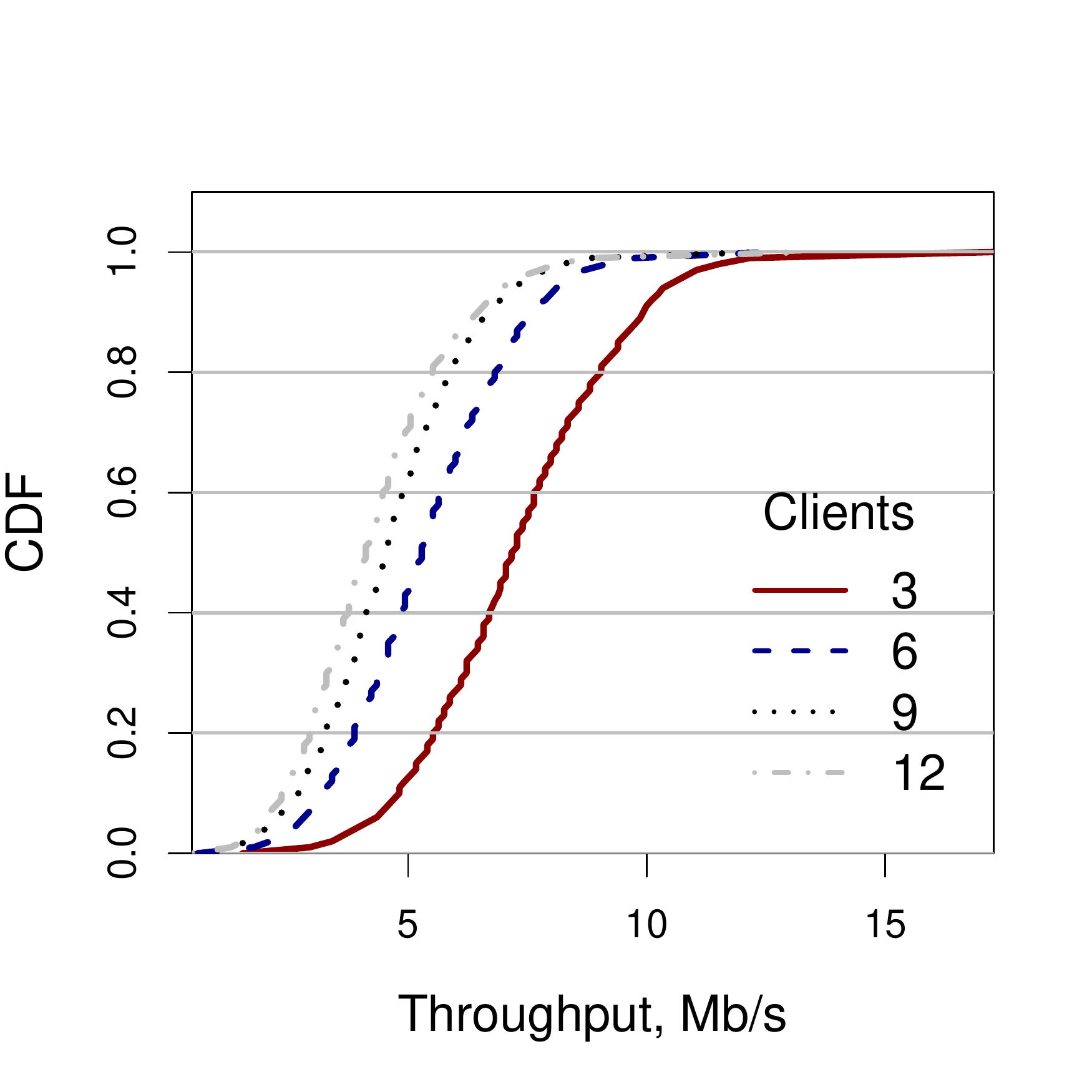}
        }
       \subfigure[802.11 standard backoff]{
            \label{fig:tput:close:enabled:std}
            \includegraphics[width=1.5in]{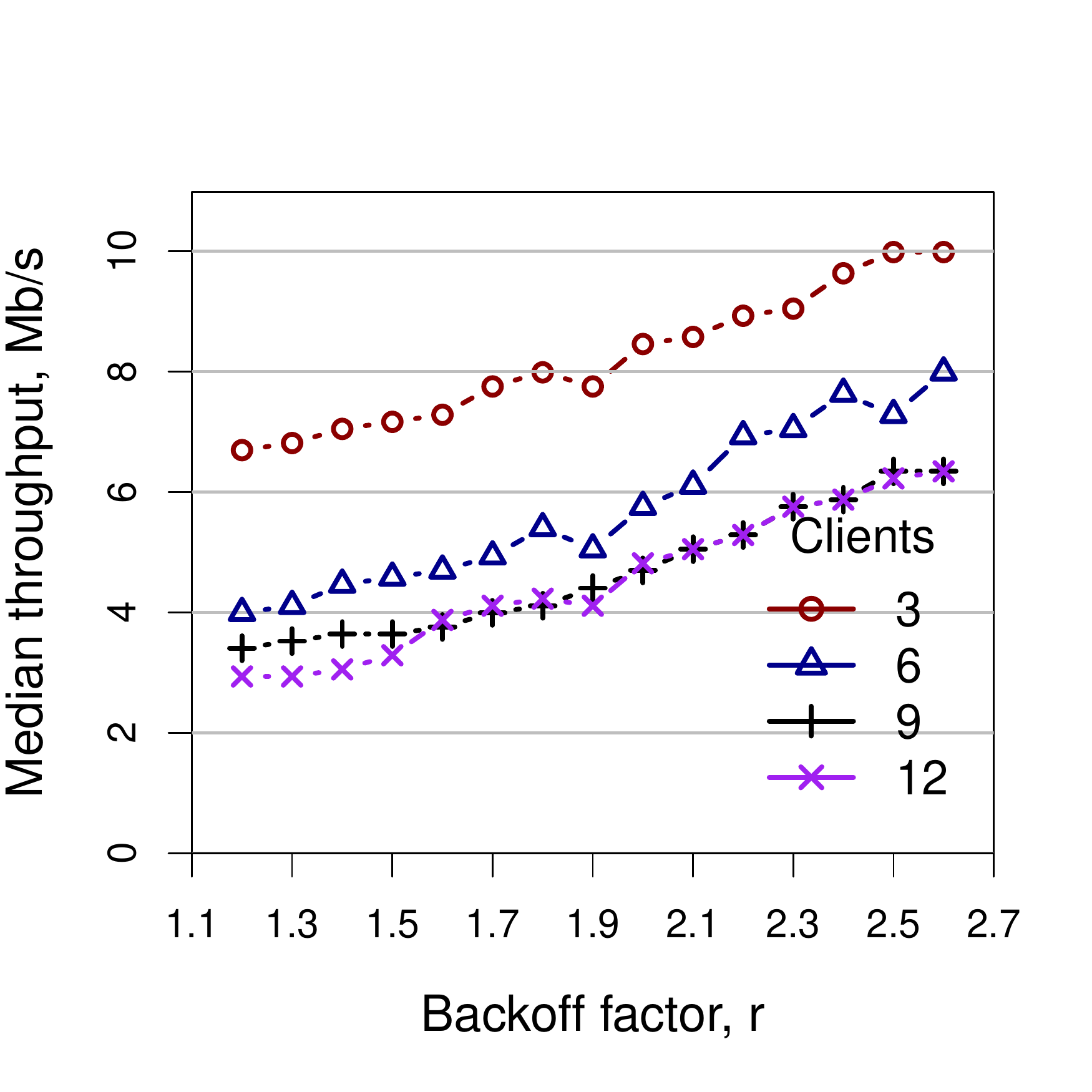}
        }
    \caption{
       Close proximity setting: Throughput
     }
    \label{fig:tput:close:enabled}
\end{figure}

Our research agenda is to observe behavior of all protocols described in Section~\ref{sec:protocols:desc} in various environments. The goal is to reveal meaningful trends by experimenting with different settings and setups. In this section we present the results for throughput (which is measured as an aggregated throughput of all stations), fairness, collision probability and delays obtained for various values of backoff factor. It is worth noting, that in our study we do not differentiate between actual packet collisions and packets losses due to other factors (such as signal fading). Instead, both events collectively constitute collisions. 

\begin{table*}[tbh]
\caption{Close proximity setting: Median throughput (Mb/s) comparison}
\begin{center}

\begin{tikzpicture}
\node (table) [inner sep=0pt] {
\begin{tabular}{c|c|c|c|c|c|c}

\backslashbox[10pt][l]{\bf Clients}{\bf Protocol} & Penalty & Rollback & Std, r=1.2 & Std, r=2.0 & Std, r=2.6 & Fixed CW \\\hline

\rowcolor{Gray}3                 & 12.57(r=1.4) & 11.16(r=1.4) &  6.69 & 8.45 & 9.98 & 7.16  \\
			   6                 & 14.33(r=1.7) & 9.16 (r=1.7) & 3.99  & 5.75 & 7.98 & 5.28  \\
\rowcolor{Gray}9                 & 13.86(r=1.8) & 9.63 (r=1.8) & 3.4   & 4.69 & 6.34 & 4.46   \\
			   12                & 14.09(r=2.0) & 10.22(r=1.9) & 2.9   & 4.8  & 6.34 & 4.11   \\ 
\end{tabular}
};
\draw [rounded corners=0.8em] (table.north west) rectangle (table.south east);
\end{tikzpicture}
\end{center}

\label{tbl:tput:comparison}
\end{table*}

We begin with the environment in which nodes are close to access point ($<1m$, Figure~\ref{fig:testbed:simple}). Next, we present results for the settings where the nodes were deployed in the environment which is typical to many office-like deployments of 802.11 networks. This experiment is followed by the scenario with two hidden stations. In all three above mentioned experiments the nodes were uploading a large file to the server.

We conclude our empirical evaluation with experiments that involve two additional traffic patterns. The first pattern we consider is when all nodes were downloading a large file. Here our intention was to observe whether the behavior of the protocols will change in comparison to the experiment with upload traffic. For the second pattern we have emulated a mixed traffic pattern with the presence of low rate voice-over-IP (VoIP) like flow and multiple (upload) TCP streams. The idea was to observe the impact of bulky streams on less demanding, but delay sensitive flows. In this experiment we were mainly interested in delay characteristics and how would suggested protocols compare with the behavior of standard protocol.

But before we dive into the discussion of the results, we define a metric we have used to measure fairness throughout the paper. Intuitively, fairness is
the ability of all transmitting stations to share the channel bandwidth roughly equally in a given period of time.
To formalize the above notion, we have used \textit{Jain's fairness index}~\cite{jain:fair} with a sliding normalized window of varying sizes. In other words, if $N$ is the number of stations generating traffic, we have computed Jain's index for windows $w \in \{N, 2N, 3N, \ldots, xN\}$. Here window size is expressed in number of packets and the lower it is the shorter term fairness is considered. More formally, if $\tau_j$ is the fraction of successful transmissions for station $j$ in window $w$, then the fairness index is defined as:

\begin{equation}
F(w)=\frac{(\sum^{N}_{j=1}\tau_j)^2}{N\sum^{N}_{j=1}\tau_j^2}
\end{equation}

Thus $F(w)=1$ is an indication of perfect fairness, and $F(w)=1/N$ is an indication of total unfairness.

\subsection{Close proximity setting}
\label{sec:perfect}

We begin with the experiment in which nodes are placed close to access point ($<1m$, Figure~\ref{fig:testbed:simple}). Although this environment can be thought as being idealized, we found that even in such a simple setting we had several sources of interference. For example, metal a rack located next to the test-bed was most likely distorting the signal. This setting emulates real life deployments in tight space areas, and therefore we find the results obtained in it interesting and relevant.  In the proceeding paragraphs we discuss the results for the following performance metrics: throughput, fairness and collision probability.

The results for the experiments when rate control was enabled are presented in Figure~\ref{fig:tput:close:enabled}. For each backoff factor we calculated median throughput across all throughput values for each bin. The exception is the plot with fixed contention window (Figure~\ref{fig:tput:close:enabled:fixed}) in which we plot the whole CDFs. We have also calculated mean throughput values only to find that they are close to median values shown in the plots, thus presenting them here is redundant.

Furthermore, in Table~\ref{tbl:tput:comparison} we highlight the above results using median throughput for most interesting configurations. Note that for penalty and rollback backoff protocols we use such values of $r$ which correspond to the maximum values of throughput. In other words, we use the peak values of $r$ according to figures. With this respect there are several interesting observations that we can make.

\begin{figure}[!bht]
	\centering
        \subfigure[Backoff with penalty]{
            \label{fig:fairness:close:gt}
            \includegraphics[width=1.6in]{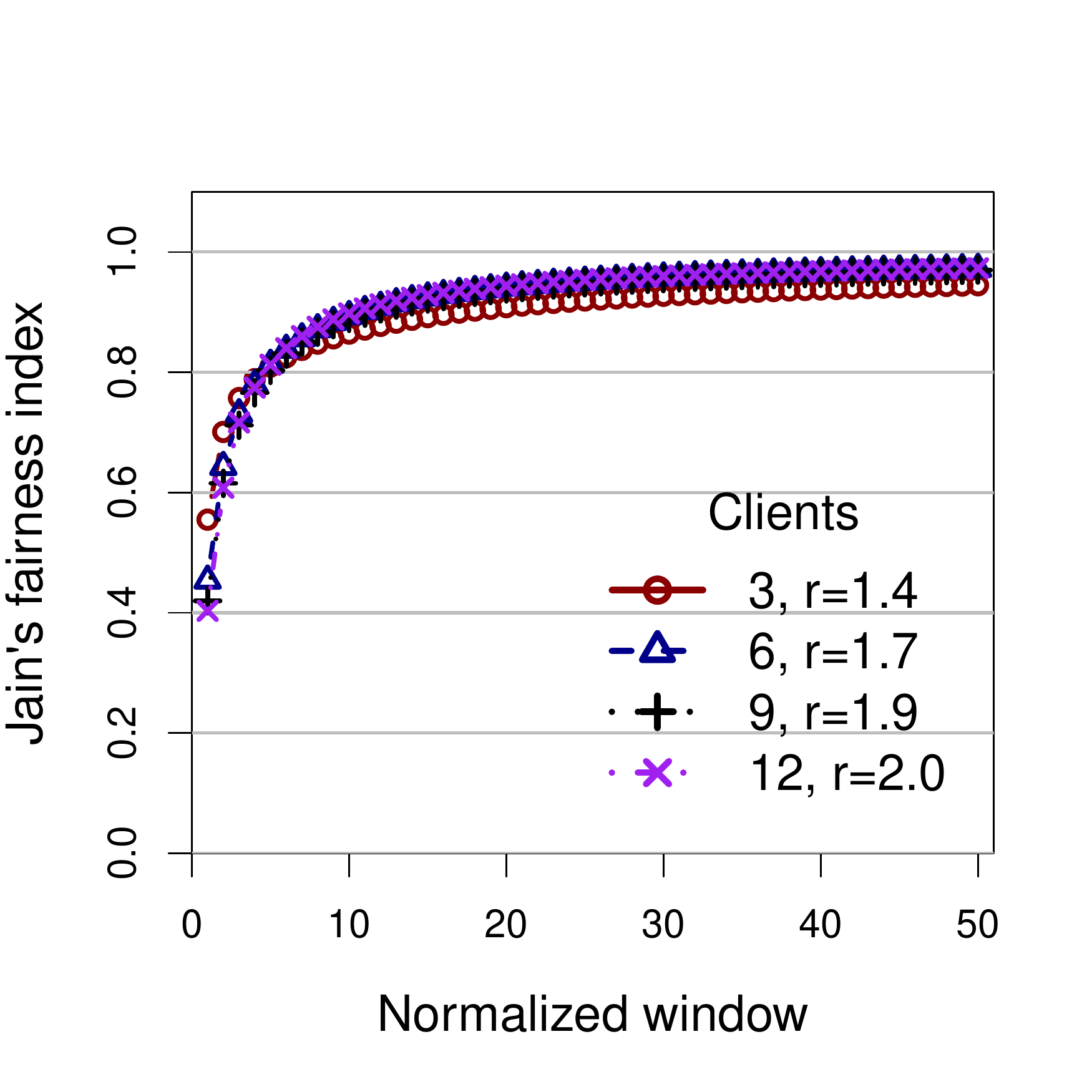}
        }
		\subfigure[Rollback backoff]{
           \label{fig:fairness:close:roll}
           \includegraphics[width=1.6in]{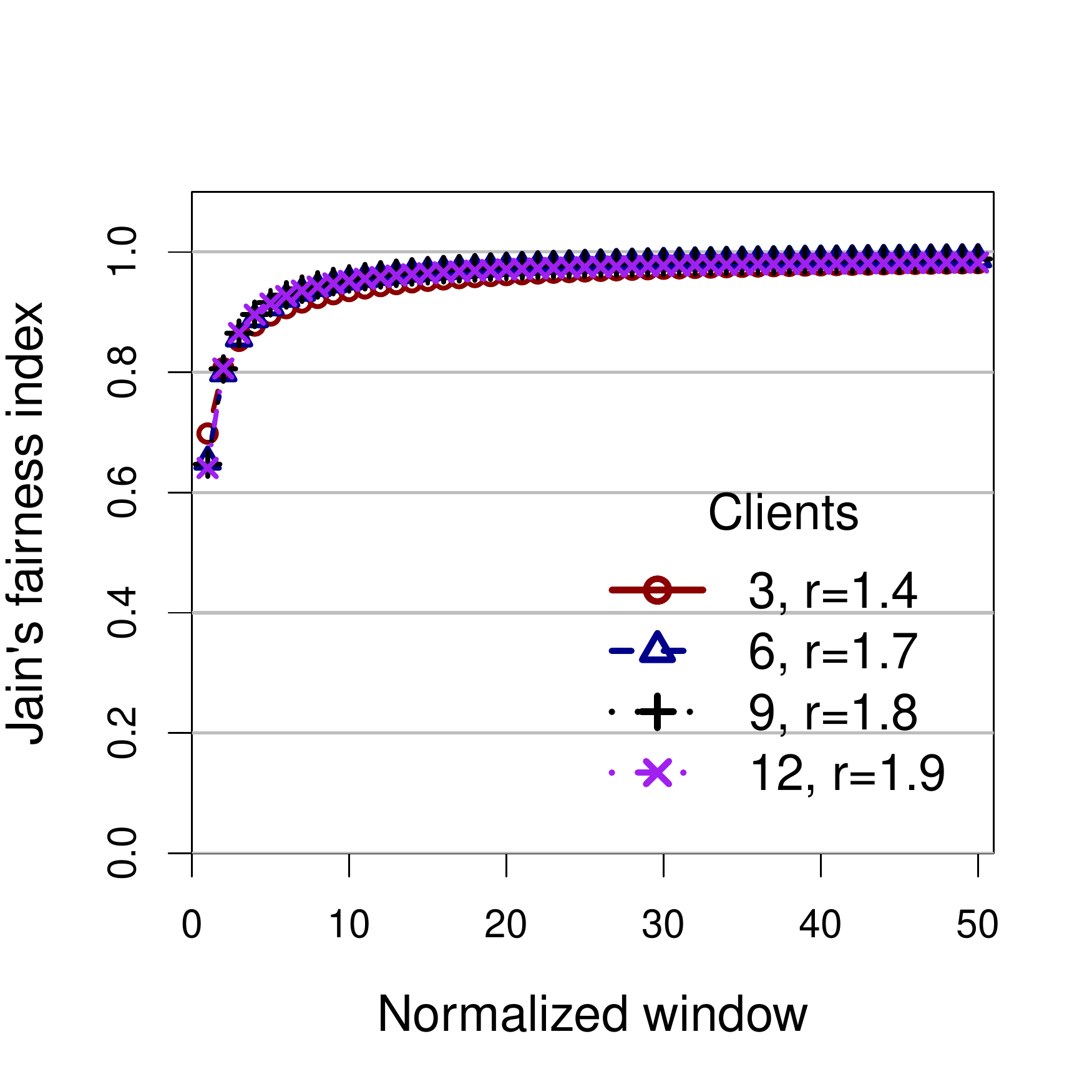}
        }
		\subfigure[Backoff with fixed contention window]{
            \label{fig:fairness:close:fixed}
            \includegraphics[width=1.6in]{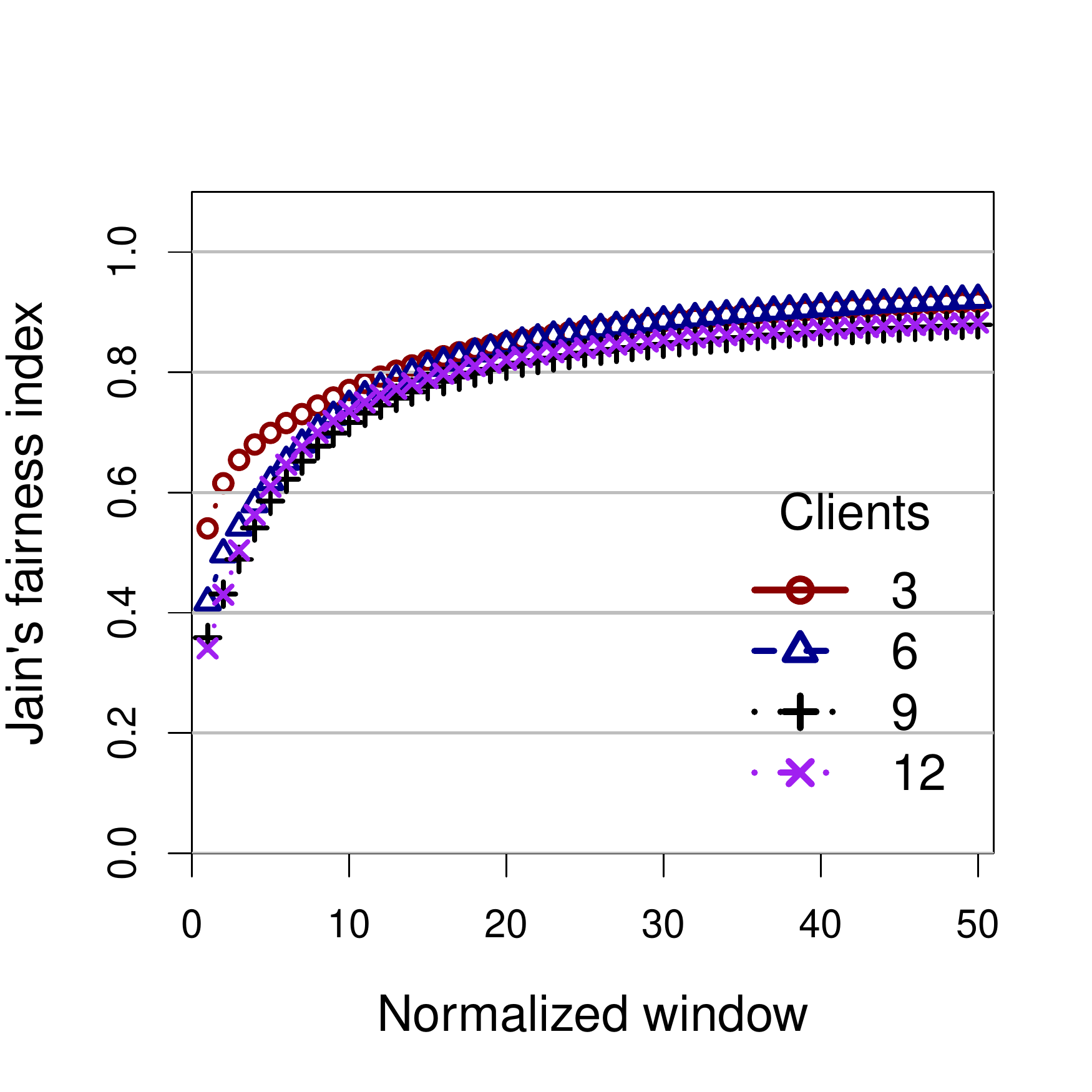}
        }
       \subfigure[802.11 standard backoff. Backoff factors $1.2$, $2.0$ and $2.6$]{
            \label{fig:fairness:close:std}
            \includegraphics[width=1.6in]{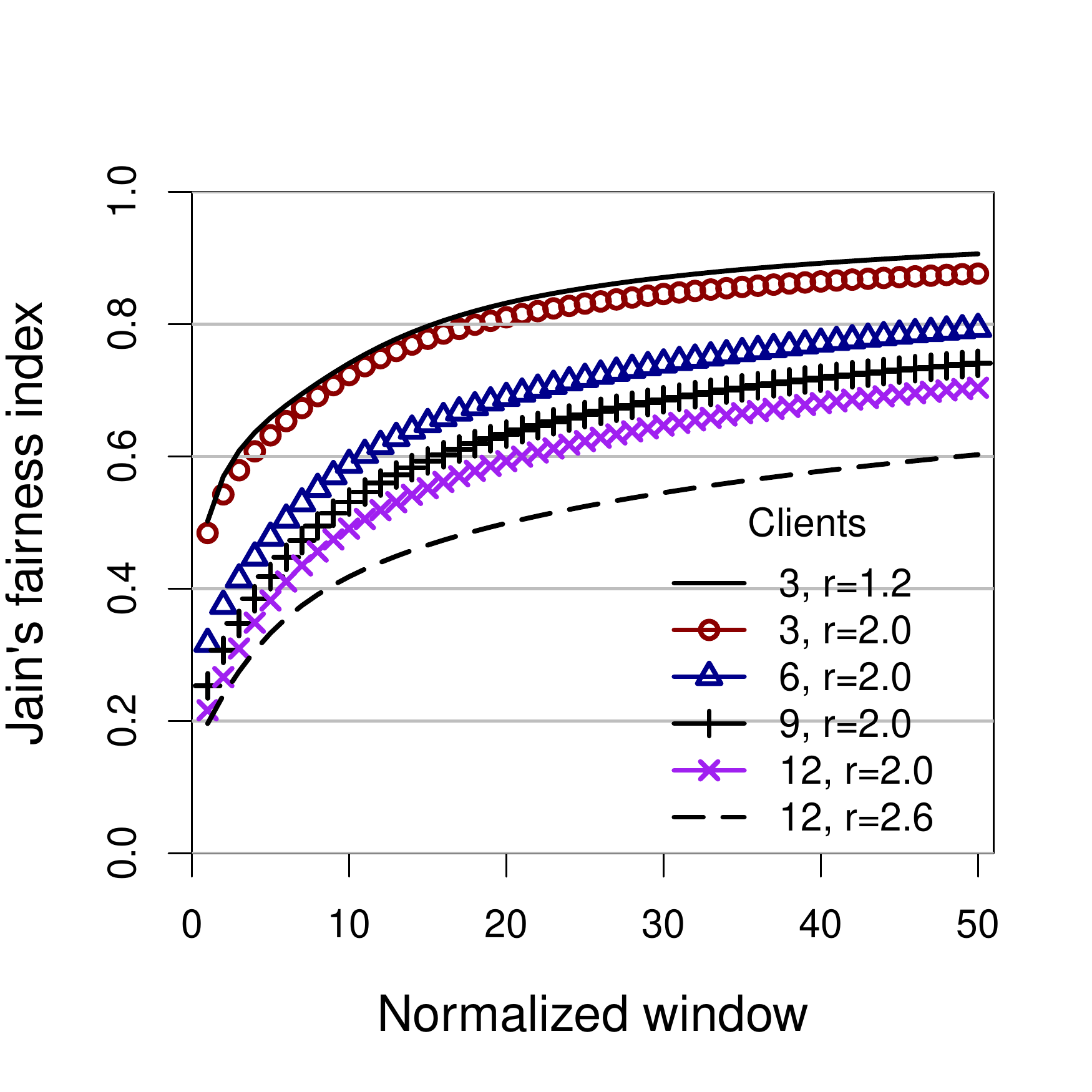}
        }
    \caption{
       Close proximity setting: Fairness  
     }
    \label{fig:fairness:close}
\end{figure}

First, for small values of $r$ (\eg $1.2$) standard 802.11 backoff protocol does not show good throughput. For this setting, the fairness is comparable to that observed for standard backoff 
with $r=2.0$ (Figure~\ref{fig:fairness:close:std}). Second, for large values of 
$r$ (\eg $2.6$) the standard 802.11 backoff shows considerable improvement in throughput, accompanied by a unduly bad fairness. As clearly seen in Figure~\ref{fig:fairness:close:std}, for $r=2.6$ fairness is worse than for $r=2.0$. In other words, larger backoff factors allow fewer hosts to capture the channel, while other host remain silent. And ff we put this in terms of user experience, it would probably mean that only few users can complete for example download, while for rest of the users the TCP connection will likely to time out (indeed, we observed such behavior in some of our experiments). For this reason in the rest of the paper we focus only on $r=2.0$ for standard backoff protocol when comparing it with other protocols. Moreover, for standard 802.11 backoff the throughput declines as the number of clients increases. In contrast, Table~\ref{tbl:tput:comparison} shows that aggregate throughput for backoff with penalty and rollback backoff protocols (for $N\in [3, 12]$)  doesn't change significantly for different number of stations. Finally, from the values presented in Table~\ref{tbl:tput:comparison} we have calculated the average improvement of rollback backoff over standard backoff ({\small $77\%$}) and over backoff with fixed contention windows ({\small $96\%$}). Similarly, the improvement for backoff with penalty is {\small $145\%$} and {\small $172\%$} correspondingly.

\begin{figure}[!bht]
	\centering
        \subfigure[Backoff with penalty]{
            \label{fig:collision:close:penalty}
            \includegraphics[width=1.6in]{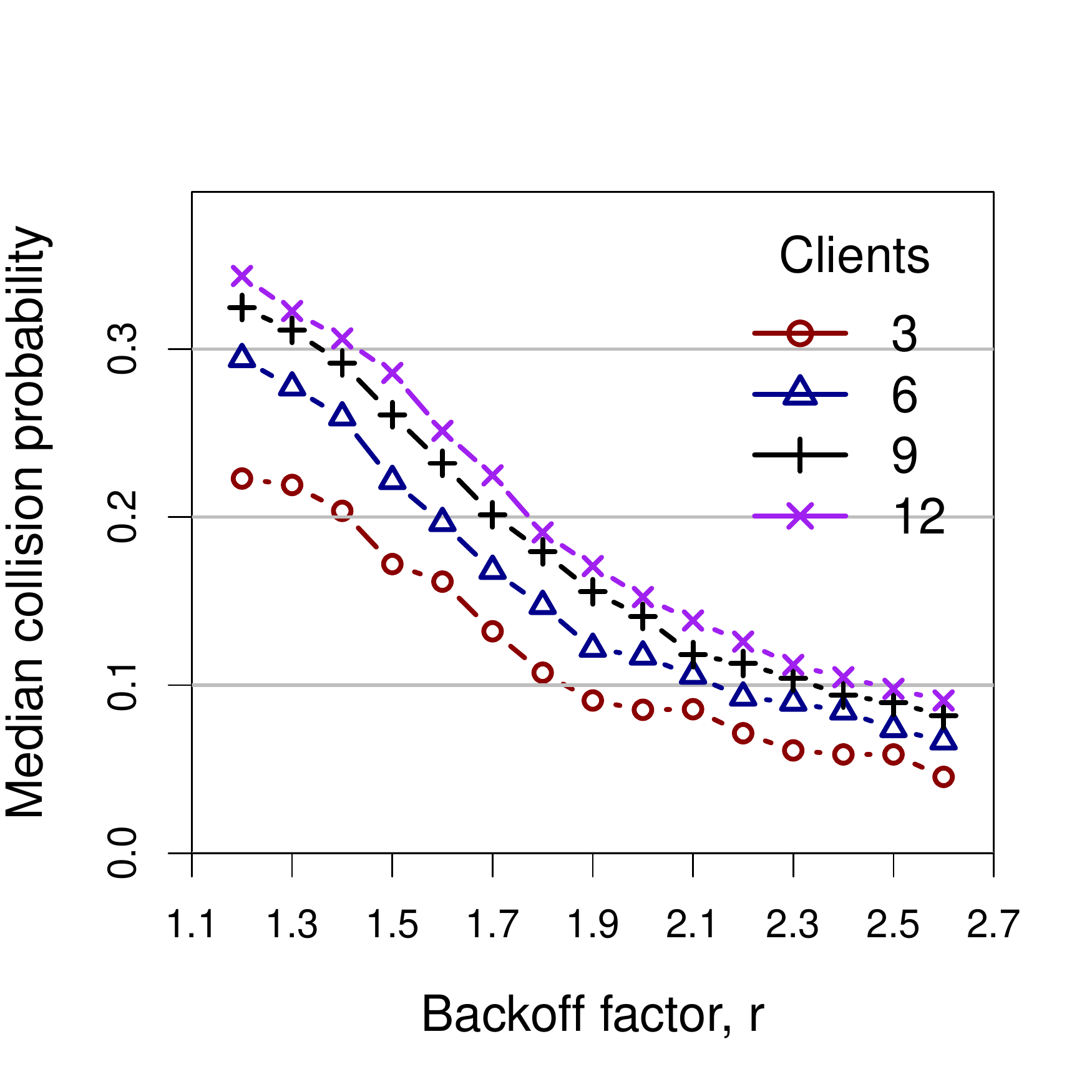}
        }
		\hspace{0.0cm}
		\subfigure[Rollback backoff]{
           \label{fig:collision:close:roll}
           \includegraphics[width=1.6in]{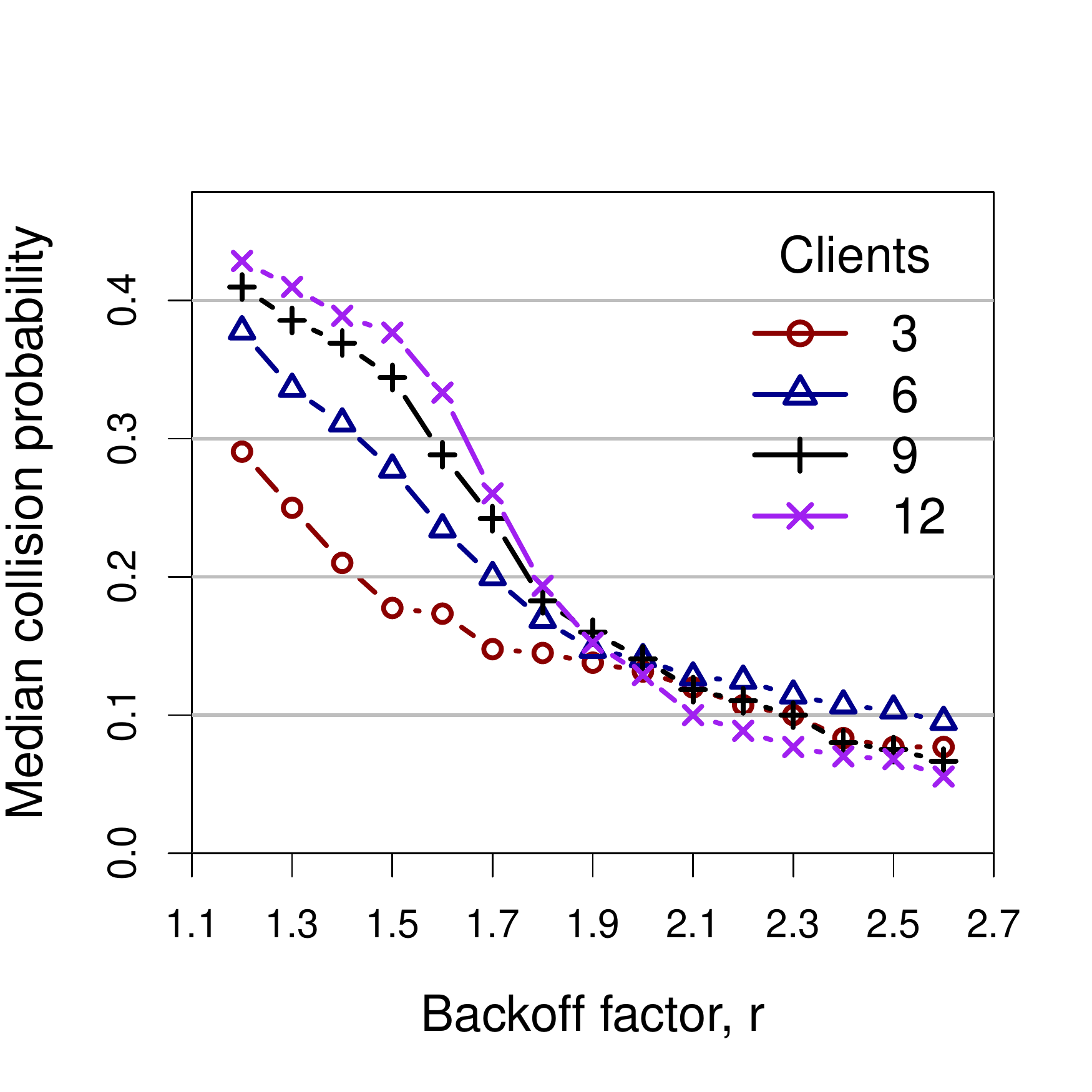}
        }
		\hspace{0.0cm}
		\subfigure[Backoff with fixed contention window]{
            \label{fig:collision:close:fixed}
            \includegraphics[width=1.6in]{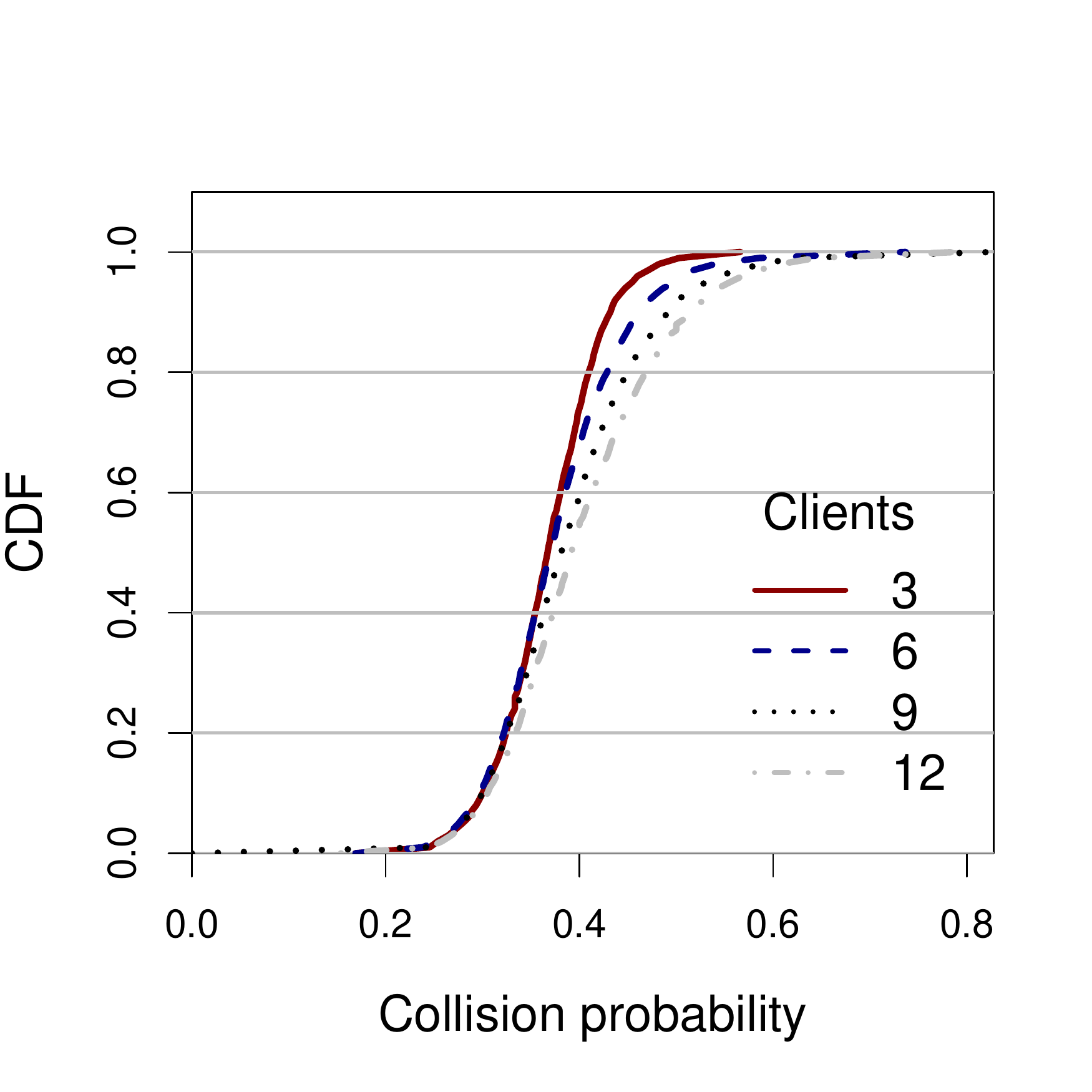}
        }
		\hspace{0.0cm}
       \subfigure[802.11 standard backoff]{
            \label{fig:collision:close:std}
            \includegraphics[width=1.6in]{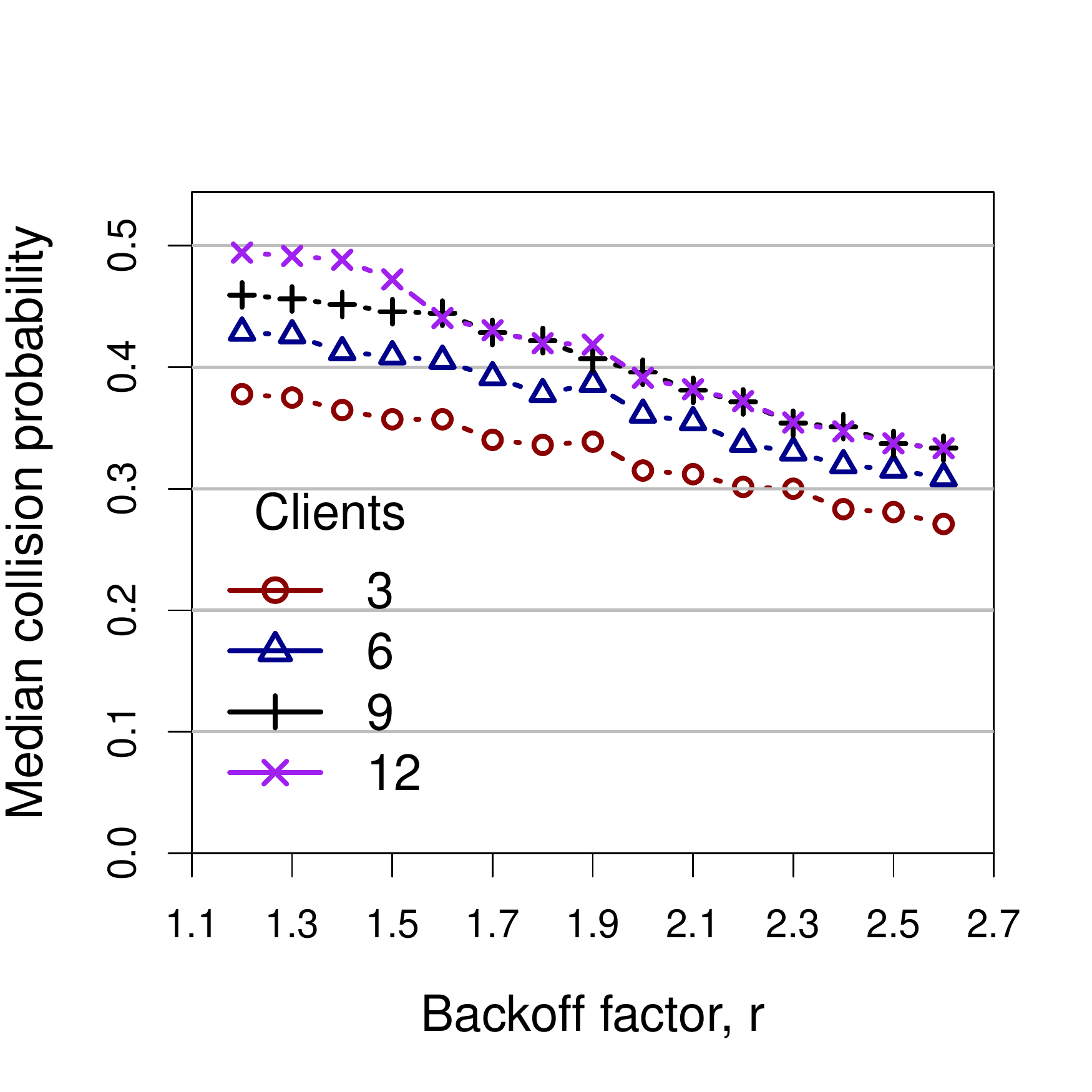}
        }
    \caption{
       Close proximity setting: Collisions
     }
    \label{fig:collision:close}
\end{figure}

For this setting we have also repeated the experiments when MAC layer rate control was disabled. We observed that for all protocols the median throughput was between $13-15 Mb/s$.
On the other hand, fairness for all protocols was similar to that when MAC layer rate control was enabled (we elaborate on fairness results in the proceeding paragraphs).
However, we think the results obtained for the settings with enabled MAC layer rate control have greater importance than the results with rate control turned off. There is extensive evidence, both theoretical and empirical, that different coding schemes deliver different bit error rates (BER) under the same signal-to-noise (SNR) level. For example, packets transmitted at $1Mb/s$ can sustain higher noise than packets transmitted at $54Mb/s$. In real networks SNR changes dynamically, which increases the role of the rate adaptation mechanism~\cite{ZhangTZWZ08, Lacage:rate:adapt}. For this reason in the remaining paragraphs we focus only on experiments with MAC layer rate control enabled. \footnote{In all our experiments wireless stations were using \texttt{minstrel} -- the default rate control algorithm in Ubuntu distributions.}


Curiously, the increase in throughput for backoff with penalty and rollback backoff does not harm fairness. The phenomena is best demonstrated in Figure~\ref{fig:fairness:close:gt} and Figure~\ref{fig:fairness:close:roll}. Thus, we can observe that both protocols achieve nearly perfect short-term fairness. On the other hand, backoff with fixed contention windows shows slightly worse results. We found that our results are somewhat different from previous work found in the literature. We see at least three reasons for the discrepancy. First, unlike other studies we do not rely solely on simulations. Second, the CW values we use in our experimental work are theoretical (see Table~\ref{tbl:ifs}), and thus, can be non-optimal in practice. Finally, it can be also that the backoff protocol with fixed contention windows does not reward the clients exposed to harsher environments (for example, in such settings the backoff with penalty can equalize the chances of successful channel access for all stations). Thus, in general we believe our results to be different due to real world effects absent in other papers. In contrast, the results for standard 802.11 backoff protocol shown in Figure~\ref{fig:fairness:close:std} were less surprising. As it is expected, standard IEEE 802.11 protocol delivers worst performance in terms of fairness.

\begin{figure}[!bht]
	\centering
    \label{fig:nonmodified:tput}
    \includegraphics[width=1.8in]{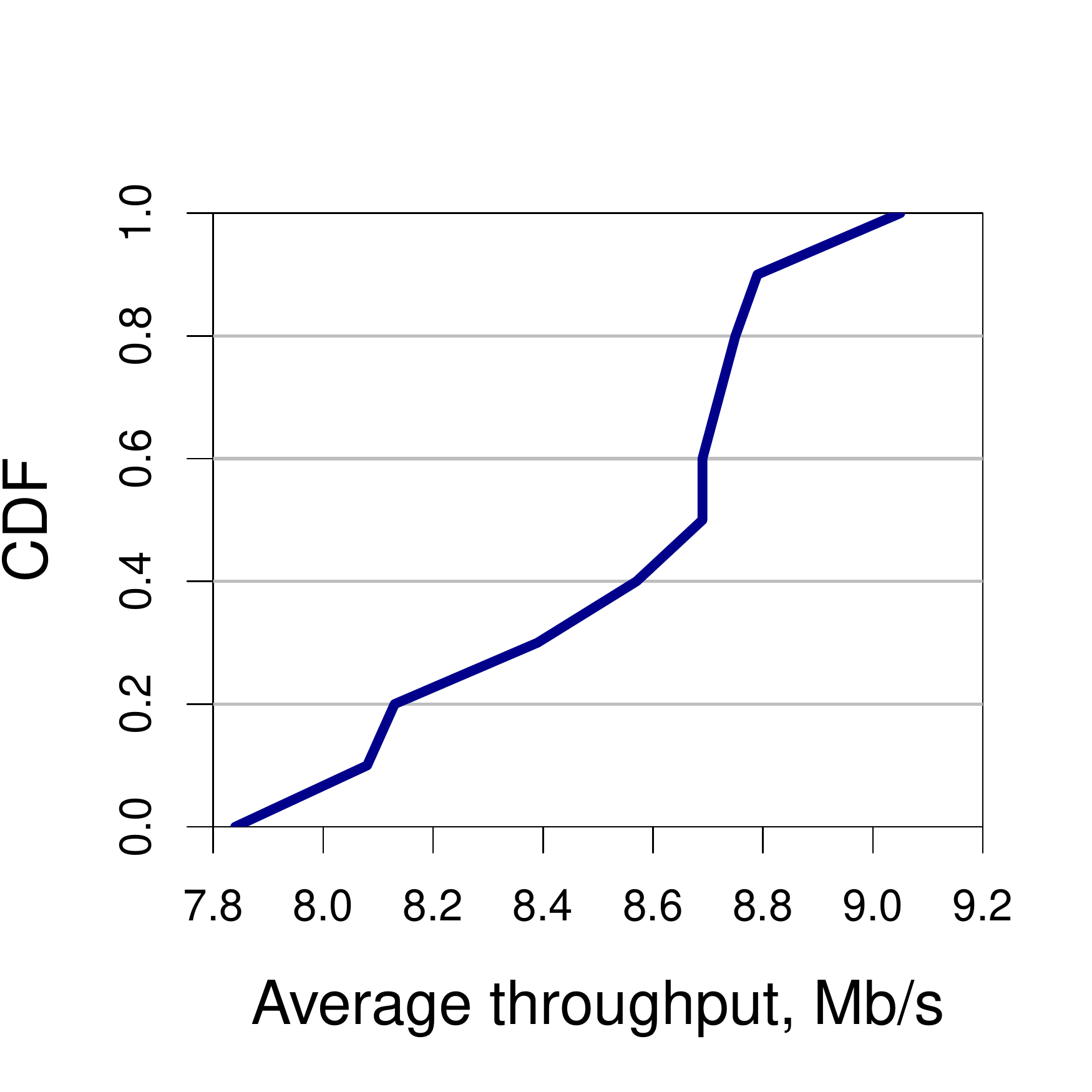}
    \caption{Aggregated upload throughput for 3 off-the-shelf laptops}
    \label{fig:validation}
\end{figure}

Finally, we turn our attention to packet collisions and their impact on system throughput. Since the MAС layer can make up to seven retries to deliver a packet, there are two ways to measure collision rate. First is to just calculate the number of packets that required more than one retry, and second is to count all the retries. Clearly, the first will yield smaller collision rates, since some packets are delivered with more than one retry. Here we report collision rates using the second approach, which explains why the numbers we obtained may seem higher than those reported in related work. 

Penalty and rollback backoff protocols (for optimal values of $r$) have sufficiently smaller fraction of collisions compared to backoff with fixed contention window and standard protocol. As can be deduced from Figure~\ref{fig:collision:close} the median collision rate in optimal throughput points varies between $0.14$ and $0.2$ for backoff with penalty and between $0.15$ and $0.21$ for rollback backoff. Similar ranges in Figures~\ref{fig:collision:close:fixed} and \ref{fig:collision:close:std} are considerably larger: from ${0.37-0.4}$ and ${0.3-0.4}$ respectively. Lower collision rate has two positive implications. First, it has a direct connection to increase in throughput; second, it improves energy efficiency of the protocols as stations spend less time in transmitting the same amount of information.

\subsection{Validation with non-modified firmware}
\label{sec:validation}


To ensure that the results we obtained for standard IEEE 802.11 protocol in 
Figure~\ref{fig:tput:close:enabled:std} are valid,
we performed an additional sanity check. We connected three off-the-shelf
laptops\footnote{In this experiment
two laptops had a preinstalled unmodified Ubuntu distribution and used \texttt{minstrel} for 
wireless rate control; the third laptop was operating under MAC OS and used a proprietary wireless rate control 
algorithm.} (which had original hardware, proprietary firmware and non-modified
drivers) to an access point which
was physically placed 1 meter away (thus, the distance was similar to
our experimental setting). The access point was
operating on an non-busy channel and itself did not contain
any custom software. On the laptops we installed an identical \texttt{HTTP}
server, which was ready to serve a 100 MB file. Lastly, using
\texttt{Wget} utility we performed concurrent downloads from all
three laptops and measured the download completion times.
We repeated the experiment for 20 times to make the results
statistically representative. From the collected data we were able to
calculate the average upload speeds which we present as a
CDF in Figure~\ref{fig:validation}. We found that the median throughput in this experiment
($8.69$Mb/s) matches closely the results of
our experimental setup with three stations running the
standard IEEE 802.11 protocol. Therefore, we concluded
that the observed throughput was normal for TCP flows over 
standard IEEE 802.11 with MAC layer rate control enabled 
and is not an artifact of our experimental setup.

\subsection{Sparse deployment}

\begin{figure*}[bht!]
	\centering
	\subfigure[Backoff with penalty]{
            \label{fig:tput:office:enabled:penalty}
            \includegraphics[width=1.5in]{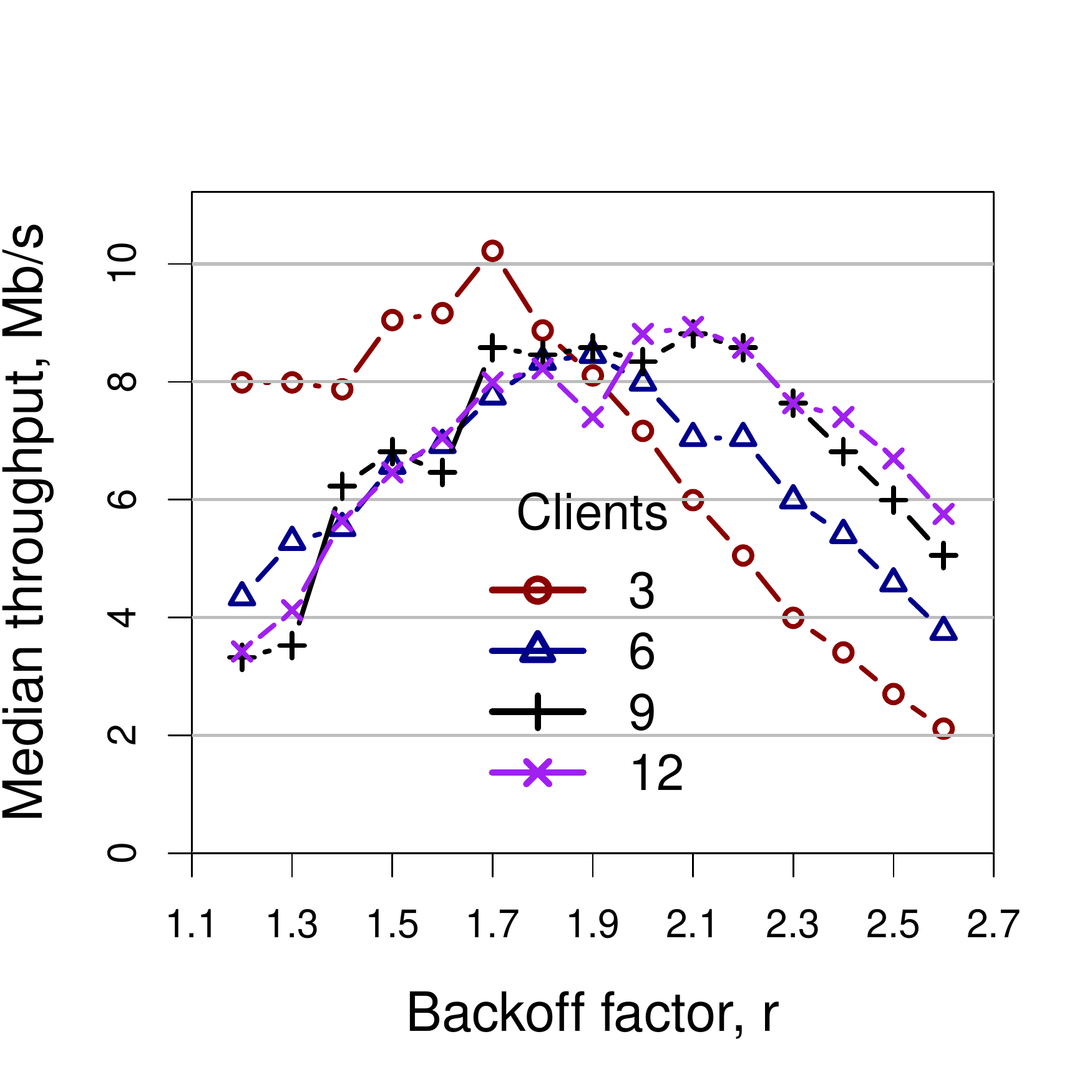}
        }
	\hspace{0.1cm}
	\subfigure[Rollback backoff]{
            \label{fig:tput:office:enabled:roll}
            \includegraphics[width=1.5in]{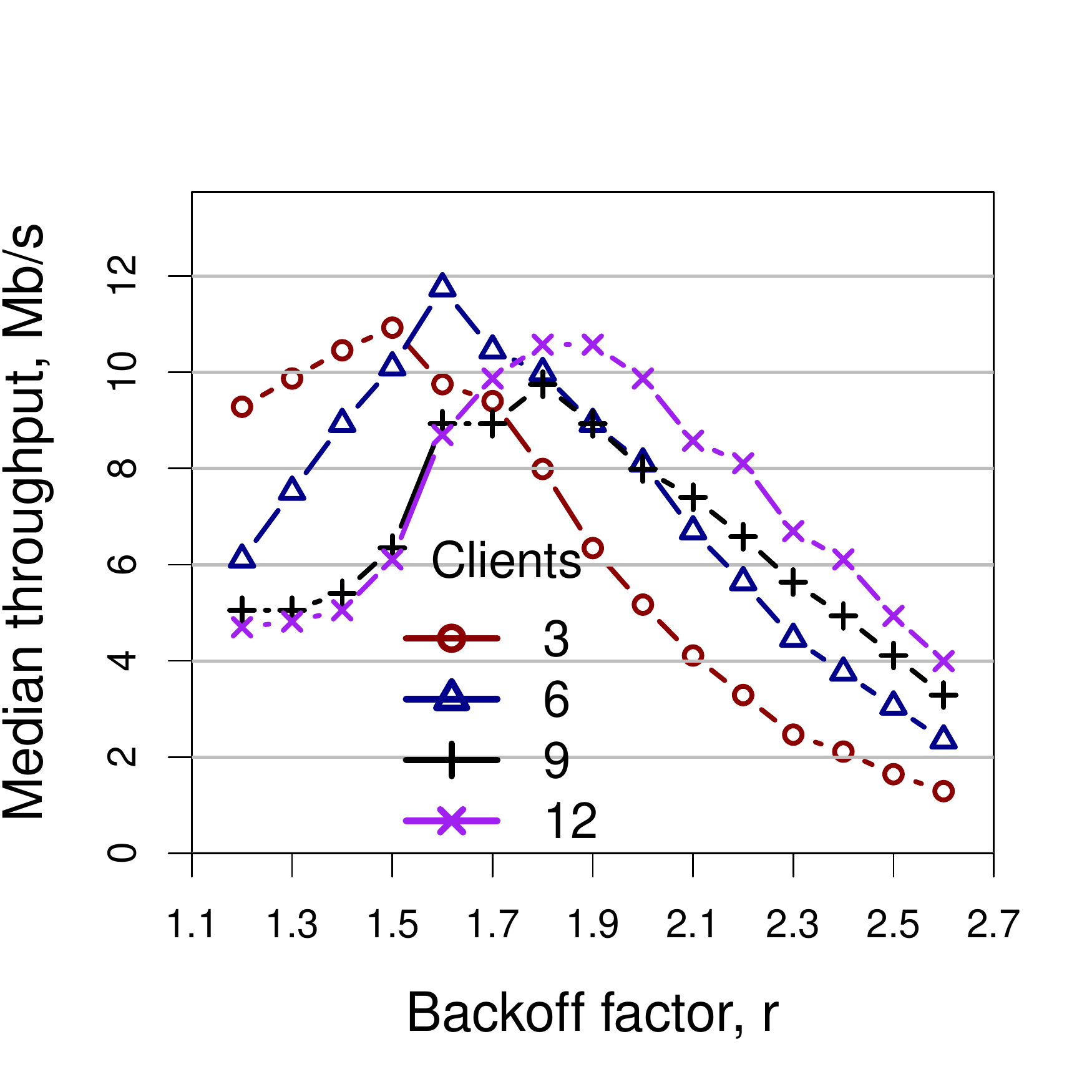}
        }
	\hspace{0.1cm}
        \subfigure[802.11 standard backoff]{
           \label{fig:tput:office:enabled:std}
           \includegraphics[width=1.5in]{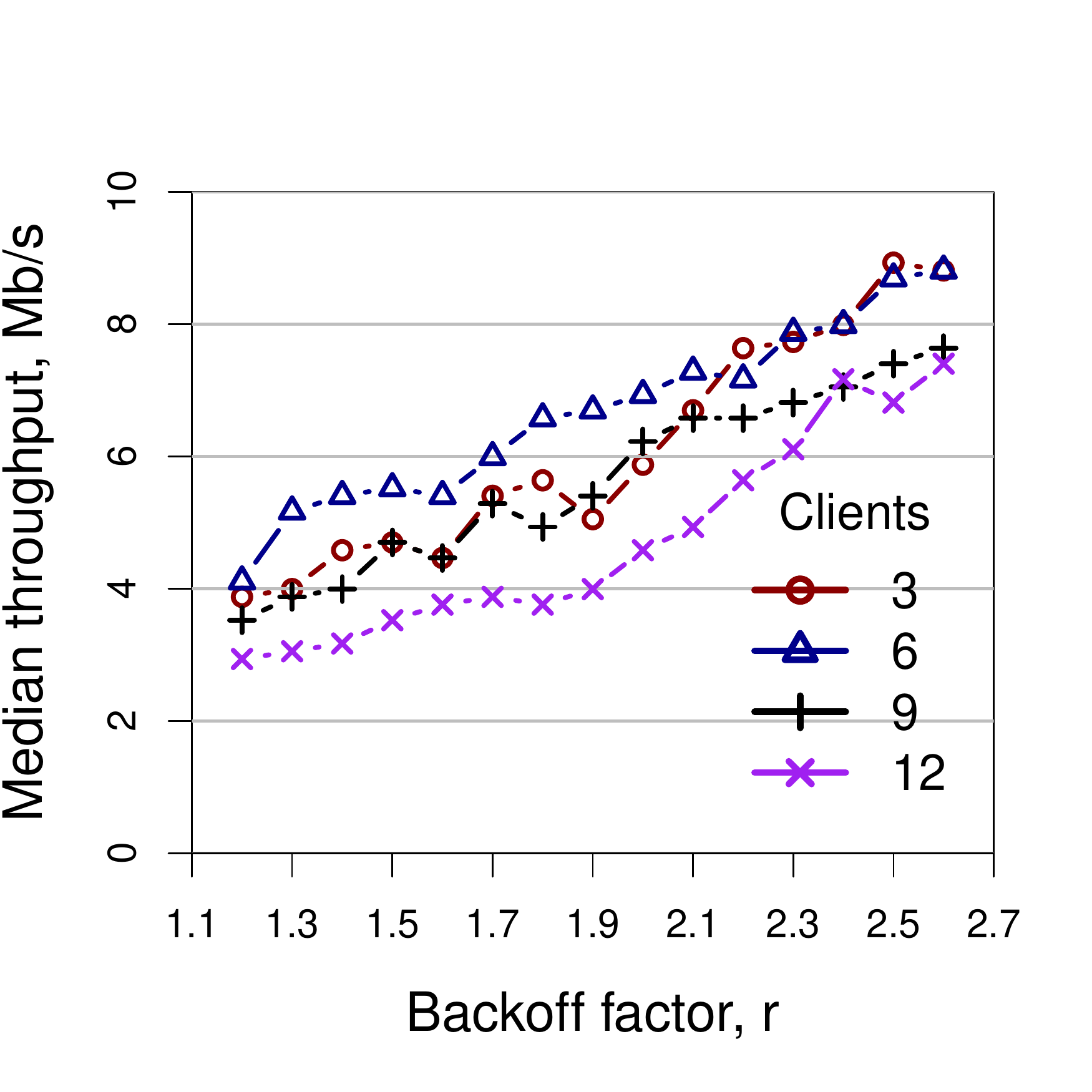}
        }
    \caption{Sparse deployment: Throughput}
    \label{fig:tput:office:enabled}
\end{figure*}

The deployment we have dealt with so far in our experimental work considers the nodes scattered around the access point no further than one meter. Although this setting allowed us to test the utility of the proposed backoff protocols, to affirm ourselves, we have conducted a set of additional experiments where nodes are placed away from each other by as much as $30$ meters. Such deployments appear in many home and office environments. Therefore, the results we obtain in this setting allow us to judge the applicability of the penalty-based protocols in realistic settings. Note, in this experiment we had several other operational wireless networks, and several networks with partially overlapping channels. Schematically the deployment looks as shown in Figure~\ref{fig:office:map}.

\begin{figure}[thp]
    \centering
    \includegraphics[width=1.8in]{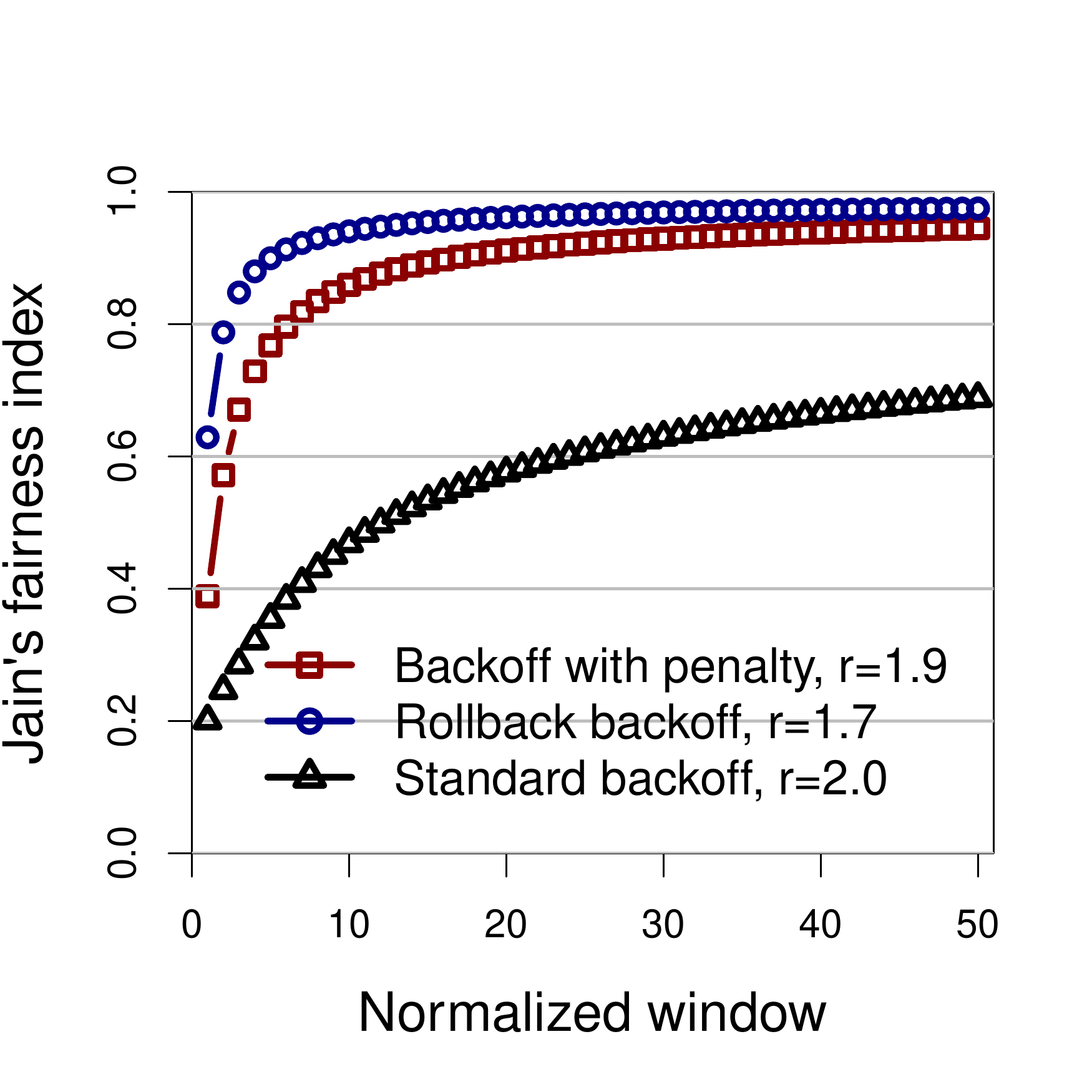}
    \caption{Sparse deployment: Fairness. 12 stations}
	\label{fig:fairness:noisy}
\end{figure}

There are several interesting observation we can make here. One such observation relates to throughput results. In general, the curves we obtain for sparse environment resemble those we presented in the previous section. This is best demonstrated by Figure~\ref{fig:tput:office:enabled}. Similarly to the experiments in close proximity environment, we have calculated the average improvement of the penalty based protocols over the standard 802.11 backoff protocol. The improvement for the rollback backoff was $>70\%$ in comparison to standard backoff. The result for the backoff with penalty was more modest: $>55\%$. Secondly, fairness of backoff with penalty and rollback backoff was again much better than in the other two protocols. Thus, in Figure~\ref{fig:fairness:noisy} we show fairness results for 12 stations only because they are almost identical to those obtained in the close proximity setting.

\subsection{Hidden stations}
\label{sec:hidden}

A hidden terminal is a serious problem for wireless networks. Briefly, it can be described as a scenario when two stations are placed far enough (or have an obstacle in between) so that they can't detect electromagnetic radiation from each other. Clearly, this breaks down the operation of DCF protocol and as such is a highly unwanted network condition. And it is exactly the reason why and how we have designed our next experiment. Regarding the goal of the experiment, here we wanted to observe whether the penalty and reward mechanisms, built-in in our backoff protocols, can solve the problem without additional mechanisms such as RTS/CTS.

To model such configuration we have placed two stations so that they could detect signals from the access point only but not from each other. To confirm this, we have configured one hidden station in the access point mode and tried to detect beacons from it on the other hidden terminal. We repeated the test in reverse direction to ensure that the property holds for the second hidden terminal. Another desirable property of this setting was the difference in signal strengths produced by the two stations. Thus, it was our intention to have significant difference in signal strengths for two stations connected to the access point. To achieve this we have done the following: first, we have placed one station inside a metal rack (see Figure~\ref{fig:testbed:simple}); and second, we have wrapped antennas of both stations with metal foil. 

\begin{figure}[!bht]
\centering
\begin{minipage}[b]{1.6in}
\centering
\includegraphics[width=1.6in]{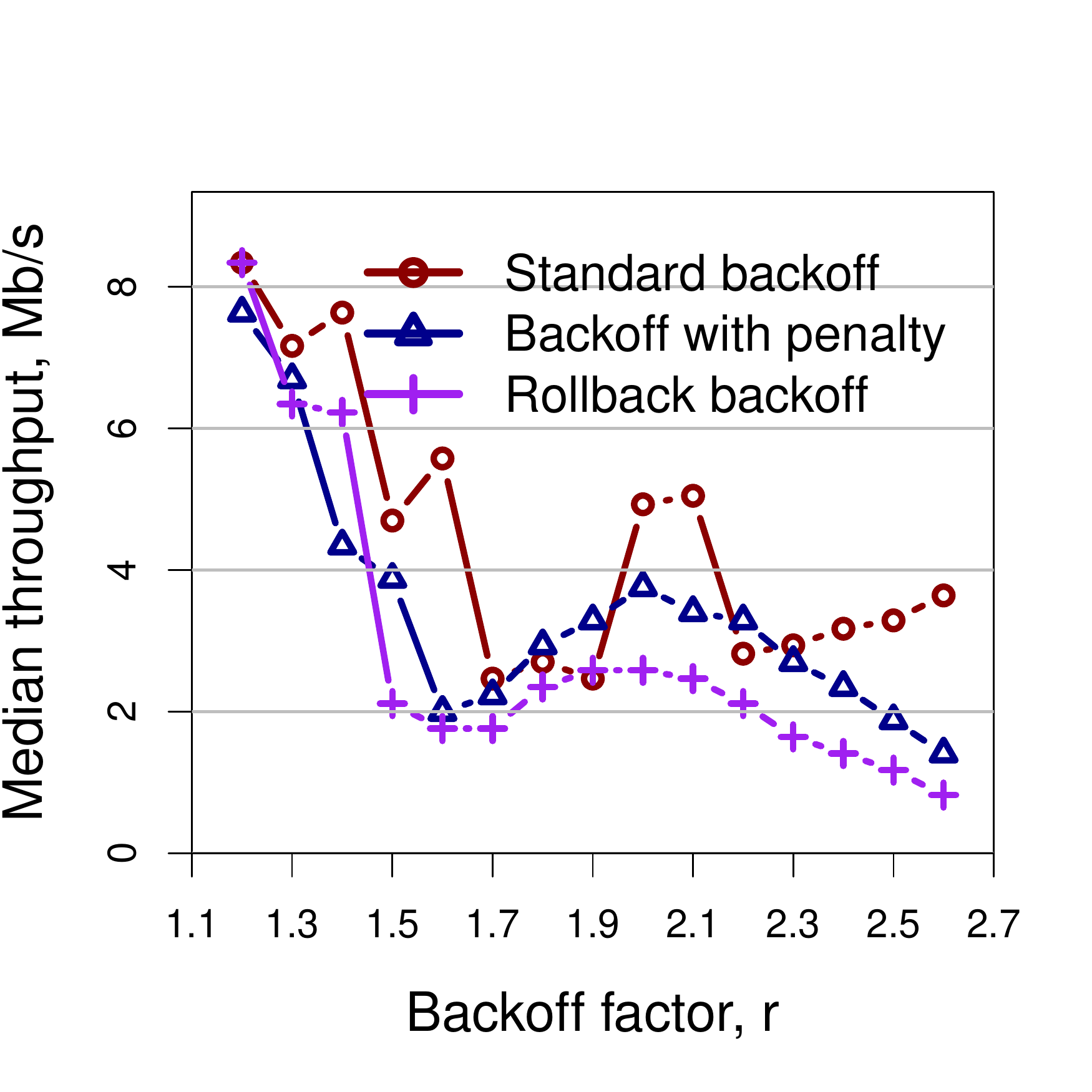}
\caption{Experiment with 2 hidden stations. Throughput}
\label{fig:hidden:tput}
\end{minipage}
\hspace{0.1cm}
\begin{minipage}[b]{1.6in}
\centering
\includegraphics[width=1.6in]{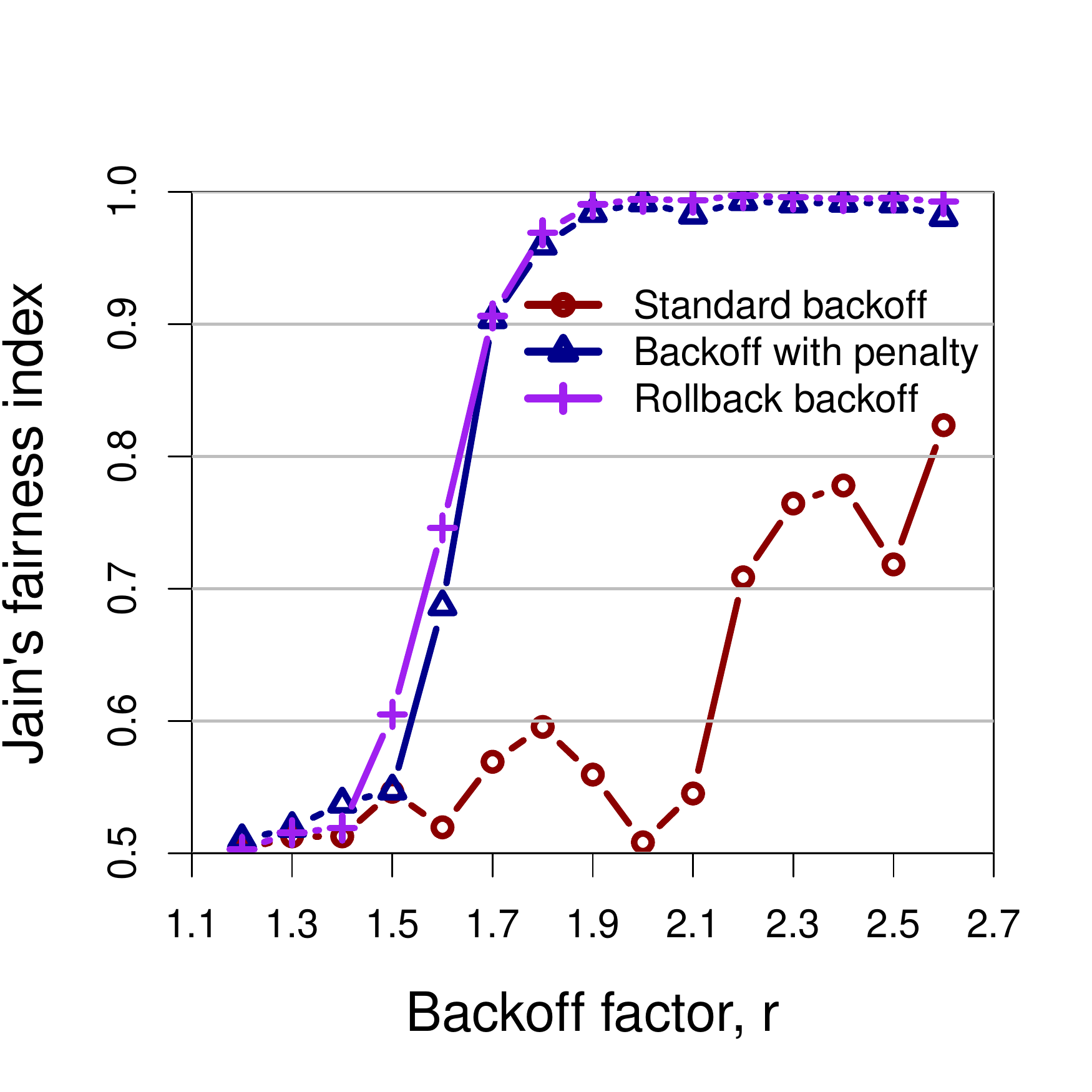}
\caption{Experiment with 2 hidden stations. Fairness}
\label{fig:hidden:fairness}
\end{minipage}
\hspace{0.1cm}
\begin{minipage}[b]{1.6in}
\centering
\includegraphics[width=1.6in]{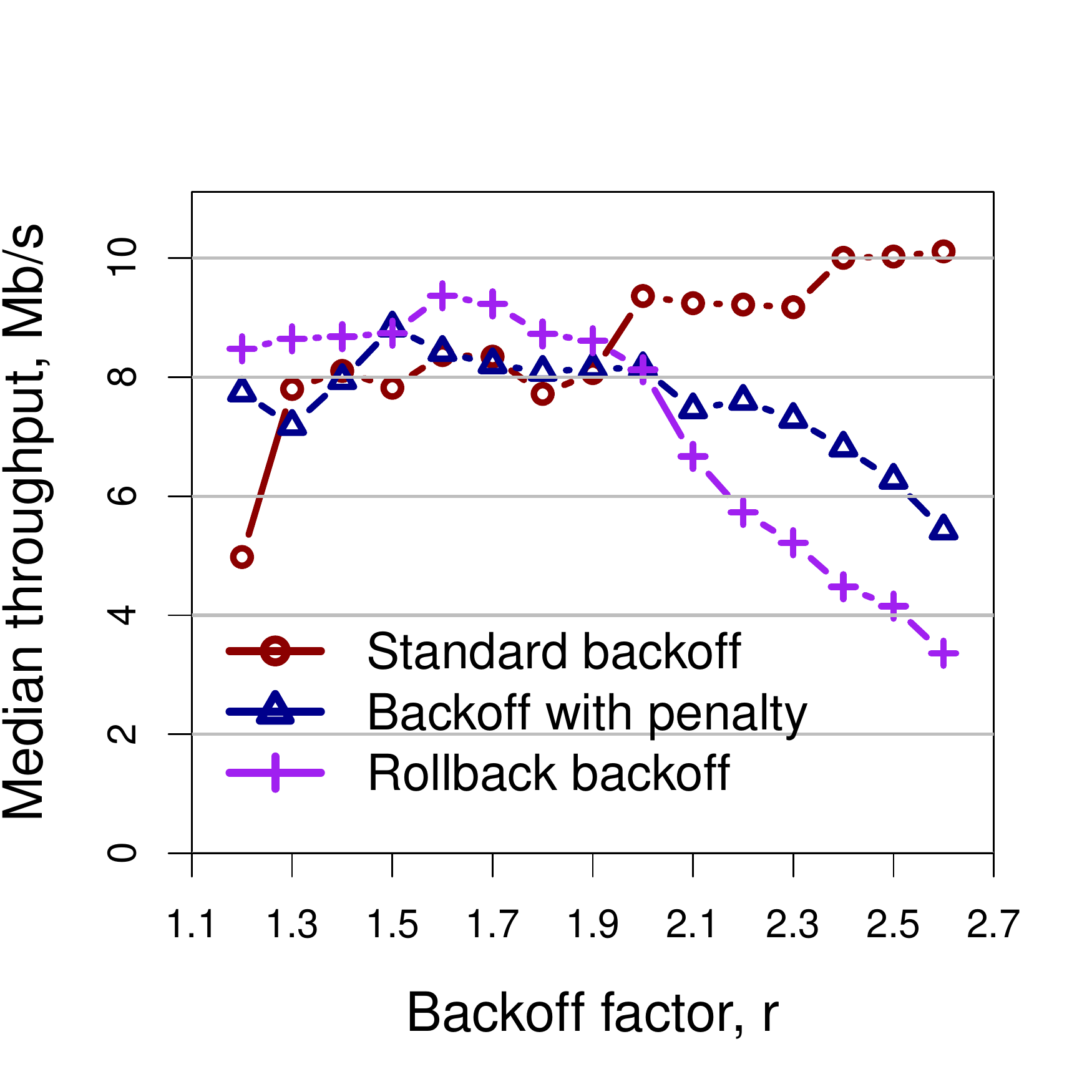}
\caption{Experiment with download traffic. Throughput}
\label{fig:download:tput}
\end{minipage}
\hspace{0.1cm}
\begin{minipage}[b]{1.6in}
\centering
\includegraphics[width=1.6in]{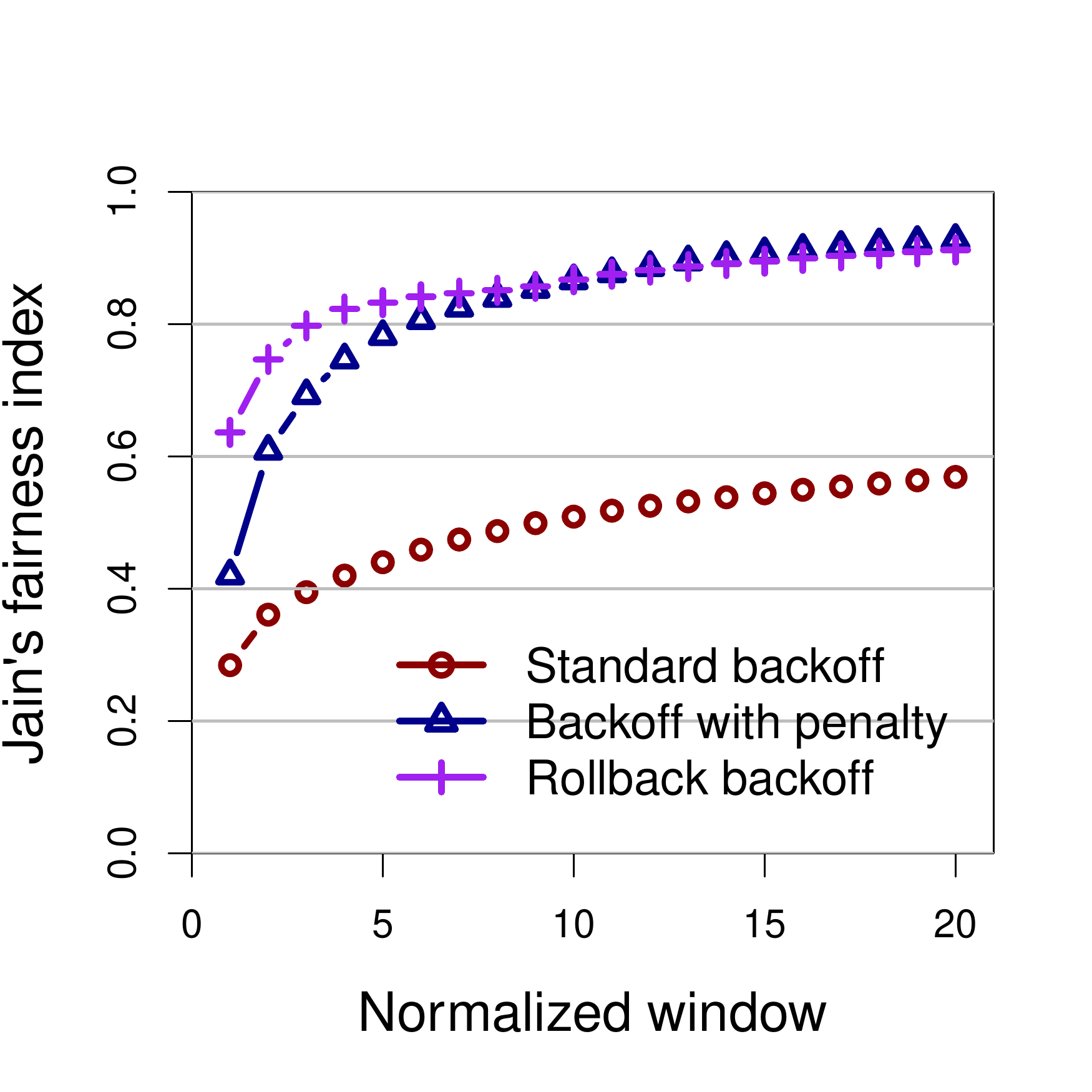}
\caption{Experiment with download traffic. Fairness}
\label{fig:download:fairness}
\end{minipage}
\end{figure}

Figure~\ref{fig:hidden:fairness} illustrates fairness with Jain's index $w=100$. There is little to no fairness for standard backoff protocol, which happens due to a single station capturing the whole channel, irrespectively of $r$. There is almost no fairness for rollback backoff when $r \le 1.4$, after which it starts growing for higher values of $r$, and eventually reaches perfect fairness. For backoff with penalty the picture is similar. It is hard to simultaneously achieve good throughput and fairness in this scenario. We further discuss this issue in Section~\ref{sect:discussion}. However, both modifications of the protocol show similar throughput results as the standard protocol, which is good.

\vspace{0.7cm}
\subsection{Download traffic pattern}
\label{sec:download}

In most wireless networks (home and public WI-FI, non-mesh networks) volume of download traffic is typically higher than volume of upload traffic. For this reason, in this section we investigate how the backoff protocols behave in a situation when download traffic pattern prevails. In this experiment the access point was following standard backoff protocol and stations were running both modified and non-modified versions of the protocols. Our intention was to understand the effect of different backoff protocols (deployed on the stations) on TCP performance. In contrast to upload scenario discussed in previous sections, in this case data packets are sent by the access point to the stations. Thus backoff protocols implemented on the stations manifest themselves in control over sending TCP acknowledgment packets. These packets are typically much smaller than data packets and due to TCP's "ACK every second" policy are approximately twice as rare.

We present the results for throughput in Figure~\ref{fig:download:tput}. Backoff with penalty and reverse backoff do show the highest throughput for optimal $r$ (in this experiment $N=7$, therefor optimal $r$, according to Table~\ref{tbl:ifs}, is $1.5$ for rollback backoff and $1.7$ for backoff with penalty correspondingly). Although, these results are comparable to those observed for the standard backoff ($r=2.0$). To our surprise, even in such configuration backoff with penalty and rollback backoff, according to Figure~\ref{fig:download:fairness}, both achieve far superior fairness than the standard protocol.

This experiment clearly demonstrates the benefit of the two modified protocols for the networks with prevailing download traffic. 

\subsection{Delay sensitive traffic}
\label{sec:delay}

Our final experiment was designed to understand the impact of our protocols on delay sensitive traffic under the presence of bulky TCP flows. We started concurrent TCP streams on $9$ clients and at the same time a single low rate UDP flow on another client. In this setting the UDP flow, limited to $80Kb/s$, was emulating a VoIP-like call. Such rate is roughly equivalent to sending $100$ byte packets every $10$ms. 

To generate UDP traffic we have implemented a simple client-server application in $C$ language. The client was periodically sending UDP packets and the server timestamping the received packets using the socket option. While there were several sources of noise in the measurement, including clock skew on the server and imperfections of $usleep(usec)$ function, these were rather insignificant so we ignored them (we have measured these imperfections with synthetic tests to observe the error).

\begin{figure*}[th]
\centering
\begin{minipage}[b]{1.6in}
\centering
\includegraphics[width=1.6in]{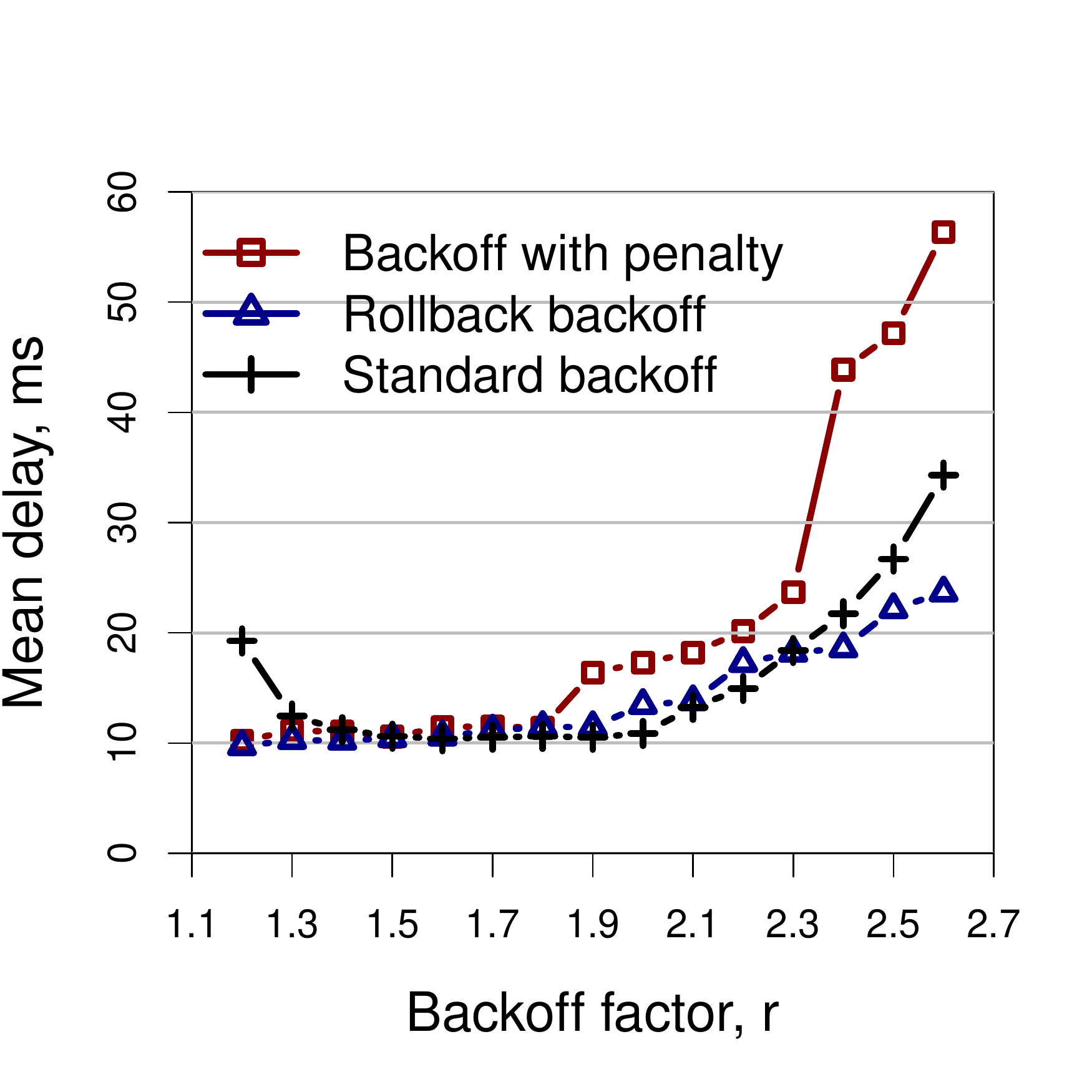}
\caption{Per packet delays observed for the experiment with mixed TCP and UDP traffic}
\label{fig:delay:udp}
\end{minipage}
\hspace{0.0cm}
\begin{minipage}[b]{1.6in}
\centering
\vspace{0pt}
\includegraphics[width=1.6in]{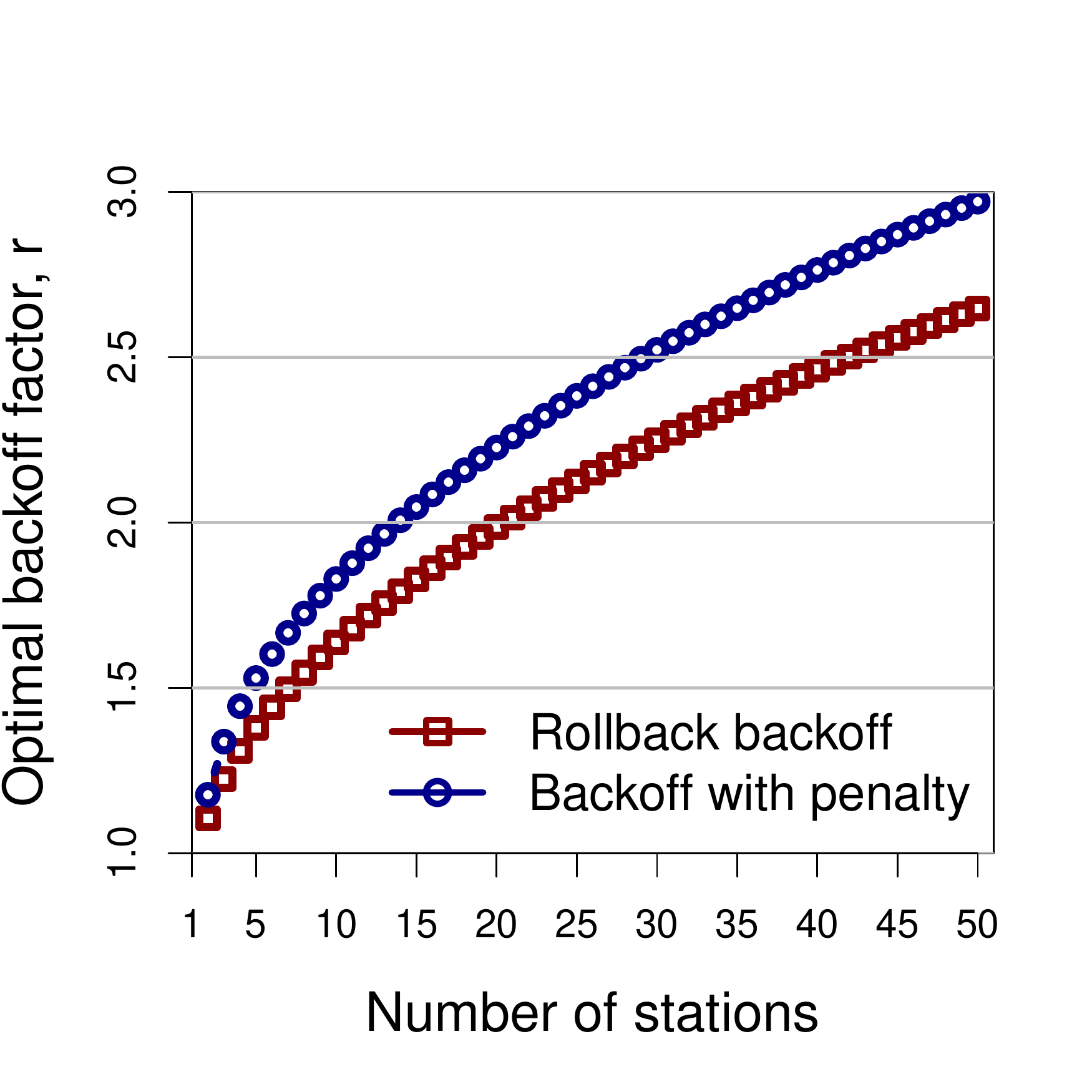}
\caption{Comparison of theoretical results for backoff with penalty and rollback backoff}
\label{fig:theoretical1}
\end{minipage}
\hspace{0.0cm}
\begin{minipage}[b]{1.5in}
\centering
\vspace{0pt}
\includegraphics[width=1.6in]{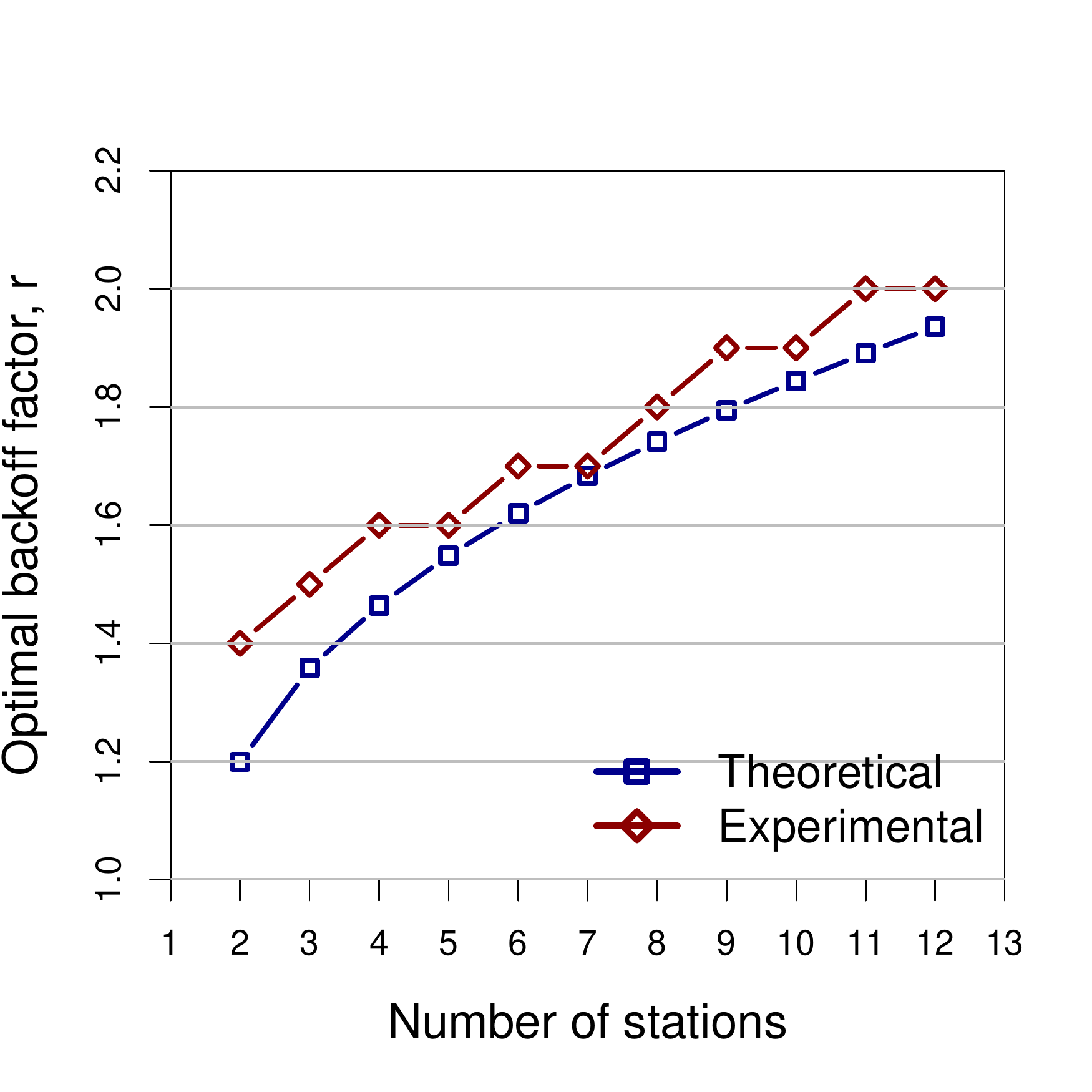}
\caption{Empirical vs. theoretical results for backoff with penalty protocol}
\label{fig:theoretical2}
\end{minipage}
\hspace{0.0cm}
\begin{minipage}[b]{1.6in}
\centering
\vspace{0pt}
\includegraphics[width=1.6in]{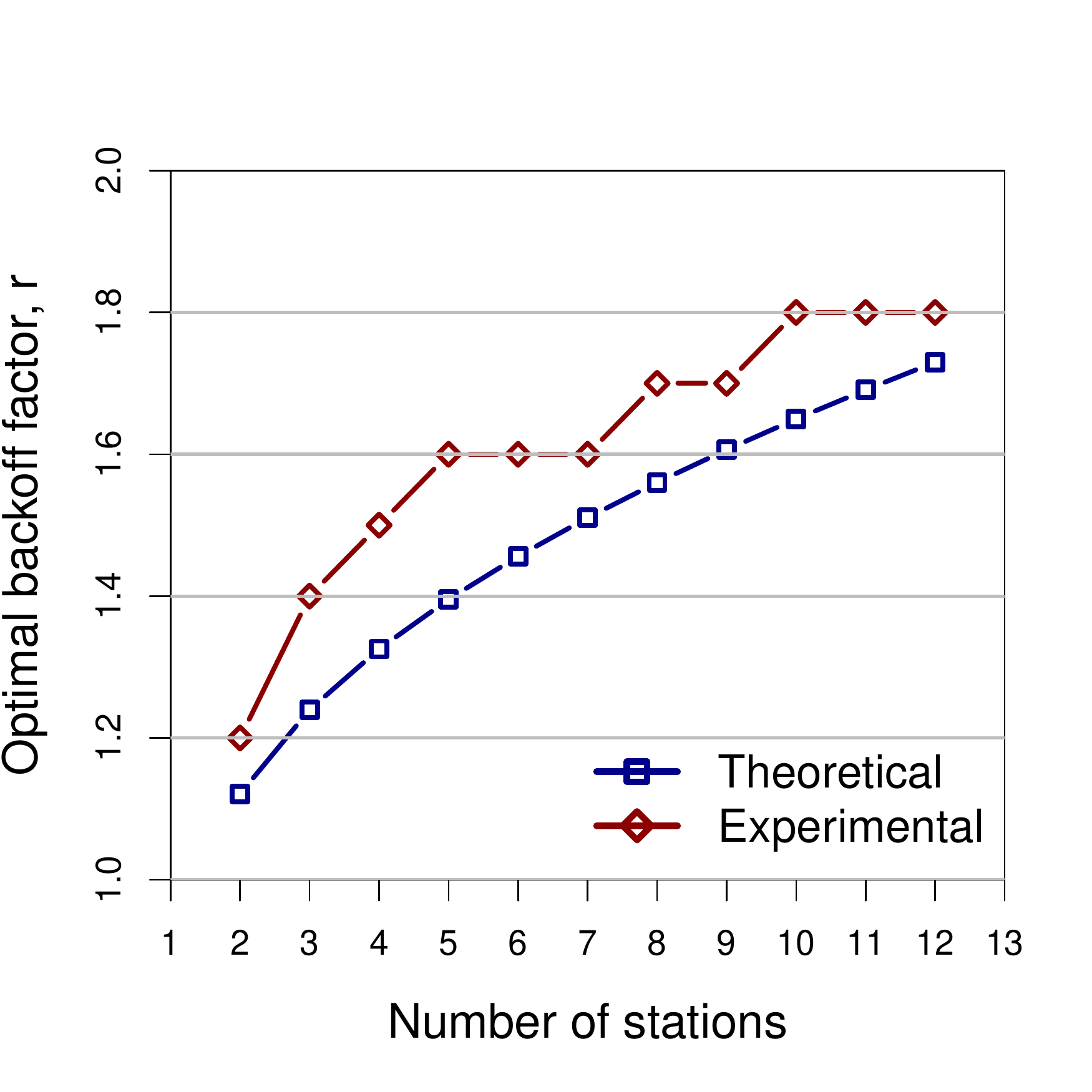}
\caption{Empirical vs. theoretical results for rollback backoff protocol}
\label{fig:theoretical3}
\end{minipage}
\end{figure*}

In Figure~\ref{fig:delay:udp} we show the results for mean delay of UDP packets. We can observe here that results for all three protocols are pretty much identical and there is little deviation from desired delay of $10$ms for small $r$. We can also notice that the delays tend to increase as $r$ gets large. We believe this is due to the backoff intervals being undesirably large, which increases the overall waiting of the queued packets. For all protocols except standard, this can become a concern when the number of stations attached to access point is large, \eg. $N > 20$ (note, $r$ grows with $N$). We should bear in mind however that for standard backoff the situation will be even worse for large $N$: $1023$ slots -- the maximum possible CW for standard backoff protocol -- will be too small for large number of stations and collisions in this case can cause more significant delays than the large CW in rollback and penalty backoff protocols for comparable $N$. Overall, we think that this problem can be tackled with better capacity planning.

\section{Simulations}
\label{sec:simulations}

\begin{figure*}[!hbt]
	\centering
	   \subfigure[Close proximity: Throughput]{
           \label{fig:simulation:tput}
           \includegraphics[width=1.8in]{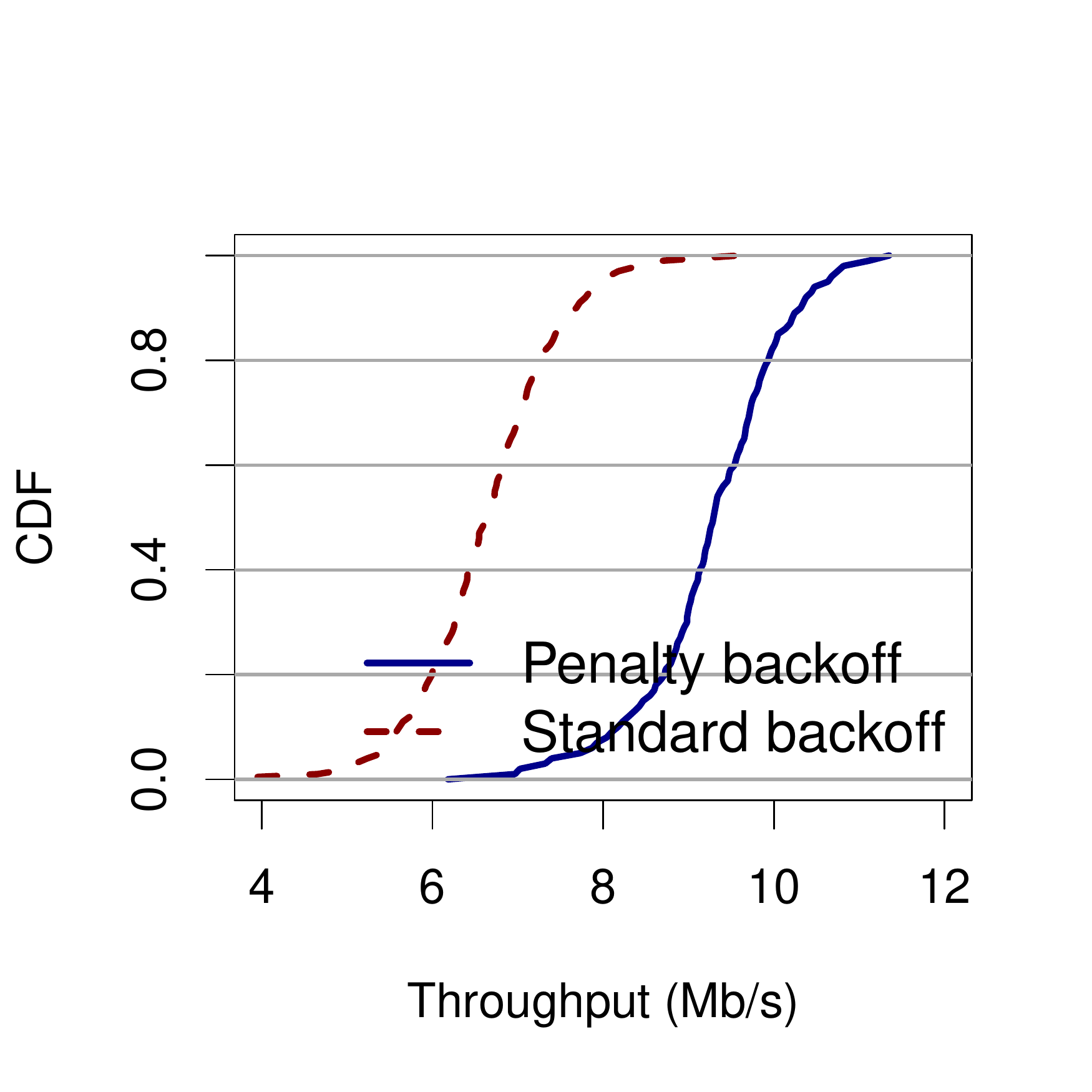}
        }
	   \subfigure[Close proximity: Fairness]{
           \label{fig:simulation:fairness}
           \includegraphics[width=1.8in]{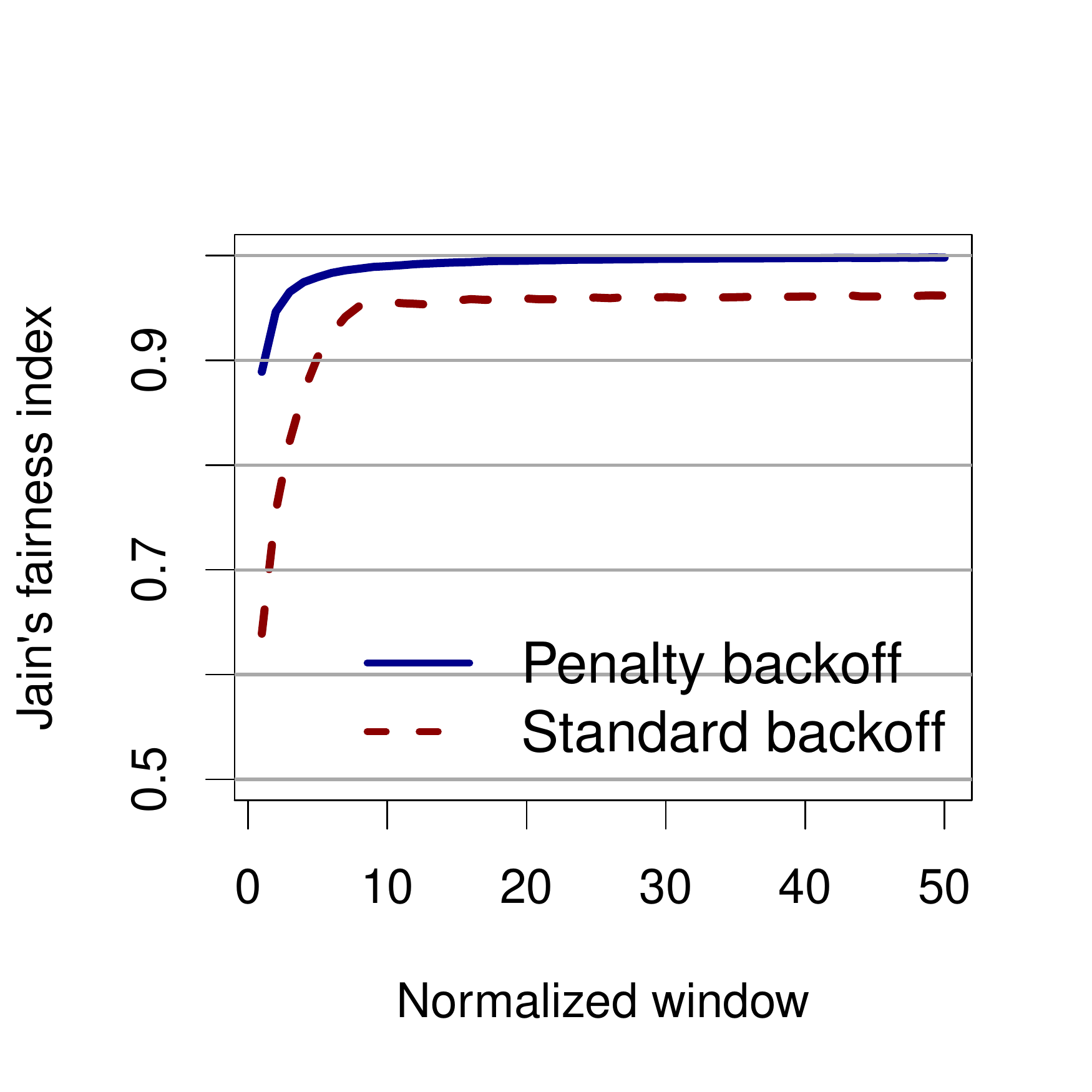}
        }
		\subfigure[Hidden stations: Throughput]{
           \label{fig:simulation:hidden}
           \includegraphics[width=1.8in]{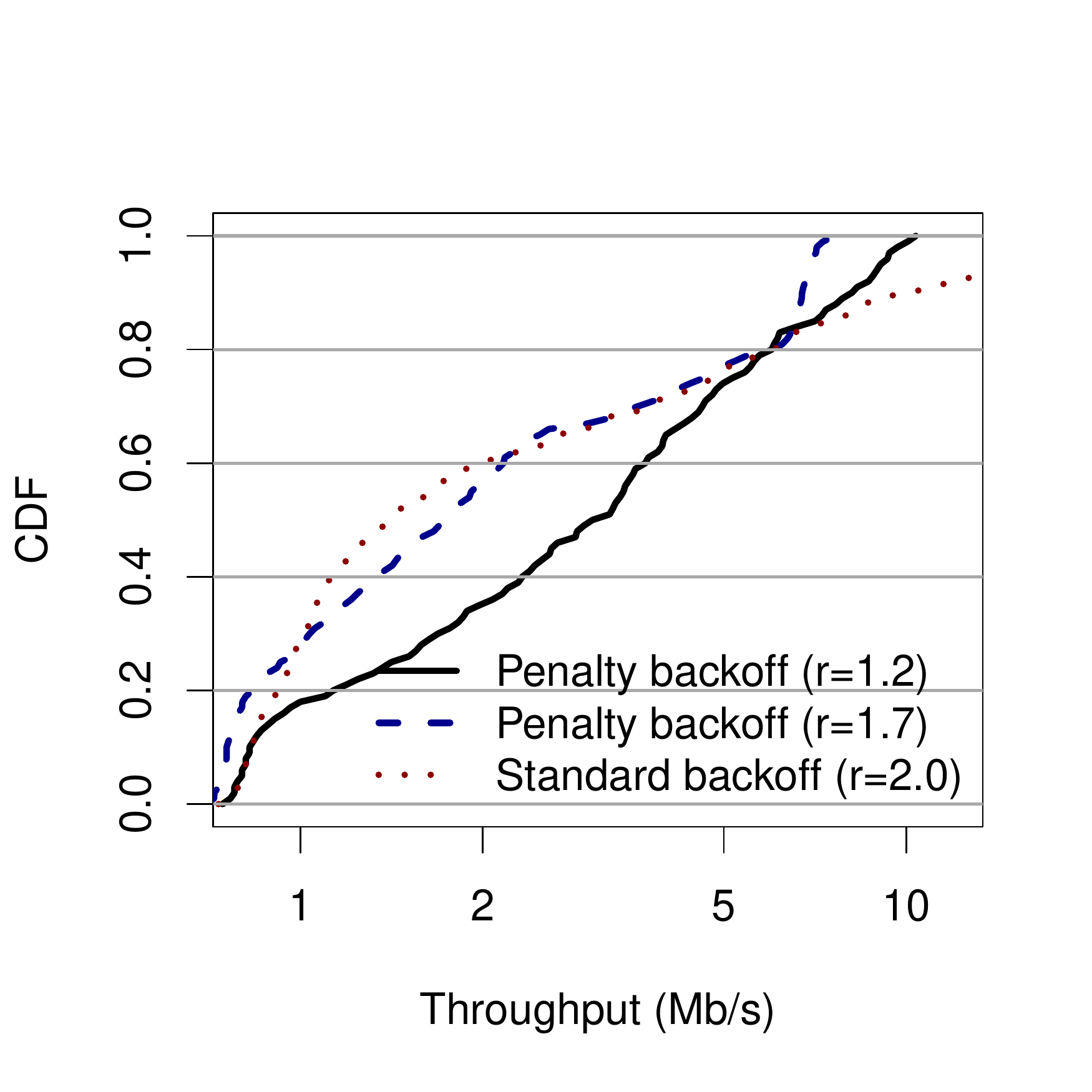}
        }
    \caption{
       Simulation results  
     }
    \label{fig:simulations}
\end{figure*}

To corroborate the results obtained in our experiments, we performed 
simulations of penalty and standard backoff protocols in NS-3~\cite{ns3:simnam} 
simulation framework. We simulated two main deployment modes described in
Section~\ref{sec:empirical}: the close proximity setting and the setting
with hidden stations. Our simulation setup included $12$ 
stations attached to an access point, with each station performing TCP upload 
to a machine attached to a wired network. Each simulation lasted 
for $600$ seconds, during which we collected the necessary data.
In our simulations all stations used 802.11g technology with 
enabled \texttt{minstrel} rate control mechanism, and RTS/CTS disabled. 
We also ensured that the queues are large enough 
to prevent packet discards due to buffer overflow, which we later
confirmed by analyzing the collected network traces. 

The analysis of the data confirmed our empirical results. 
For example, Figure~\ref{fig:simulation:tput}
and Figure~\ref{fig:simulation:fairness} show throughput 
and fairness results for the simulations with rate control enabled. 
Clearly, the trend we saw in the real life experiments of close proximity settings 
is visible here as well. 
To compare the collision rate, we measured the number of retransmitted packets. 
We found this figure obtained from simulations to be around $13\%$ and $4\%$ 
for the standard and penalty backoff protocols respectively. These numbers are
smaller than the collision rates we saw in real experiments, which is certainly
to be expected since simulation provides an idealization of many real 
mechanisms such as timers, queues, etc. Nevertheless, the overall trend 
clearly persists. 

We also repeated the experiment with two hidden stations. We configured 
the simulation parameters so as to ensure that the two uploading wireless stations
don't receive each other's signal, but can communicate with the access point.
Here, we experimented with penalty backoff with $r=1.2$ and $r=1.7$, and 
standard backoff with $r=2.0$. The results we obtain resemble those we report 
in Section~\ref{sec:hidden}. Figure~\ref{fig:simulation:hidden} demonstrates 
distributions of throughput, with median throughput being clearly better for
penalty backoff. Fairness index (with $w=50$) for standard backoff turned out
to be $0.72$, and $0.69$ and $0.93$ for penalty backoff with 
$r=1.2$ and $r=1.7$ respectively. These simulation results confirm
the validity of our real-life experiments.

\section{Analytic model}
\label{sec:analytical}
In this section we derive an analytic model describing behavior of wireless networks with rollback and penalty backoff protocols. We consider the model important for two reasons. First is to solidify and corroborate our empirical results---as we will see later in this section the proposed model agrees well with our experimental findings. Second, the model can be used to choose an optimal value of $r$ for a given number of active stations $N$, importance of which is discussed in Section~\ref{sect:discussion}. 

{\it Relationship between expected contention window and r}. Using many lines of research~\cite{tay:capacity:model,bianchi:performance,vu:collision:model} on analysis of 802.11 networks as a starting point we find the expected value of contention window for rollback backoff protocol:

\begin{equation}
\label{eq:exp:rollback}
E[CW]=\frac{(CW_{min}-1)(1-p_c)(p_c^{k}-r^{k})}{2(1-p_c^k)(p_c-r)},
\end{equation}

where $r$ is the backoff factor, $p_c$ is the probability of collision (to be defined in Equation~\ref{eq:coll:prob}), $k=7$ is the maximum number of retransmissions and $CW_{min}=16$ is the minimum contention window.

\begin{figure}[htp]
   
    \centering
    \includegraphics[width=2in]{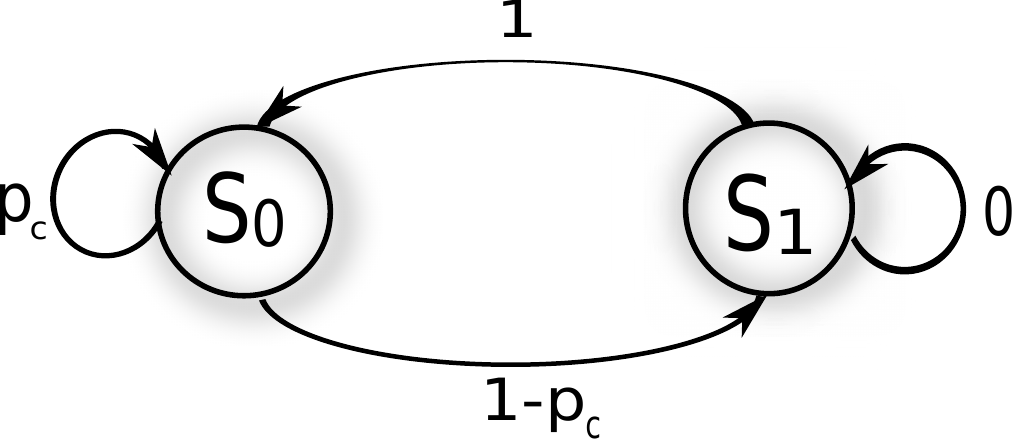}
    \caption{
       Markov chain for backoff with penalty
     }
     \label{fig:markov:chain}
\end{figure}

To model backoff with penalty we used an observation that according to Figure~\ref{fig:reverse:backoff:1} the station, besides backoff process itself, can be in two states. The first state ($S_0$) is the state in which the station fails with probability $p_c$ to transmit at the first attempt and thus follows the standard backoff protocol. The second state ($S_1$) corresponds to the case when the station succeeds to transmit a frame at the first attempt with probability $1-p_c$ and is thus forced to use the largest contention window for transmission of the next frame. We can model this process with a two-state Markov chain (Figure~\ref{fig:markov:chain}), which essentially encodes the states the backoff with penalty protocol can be in when attempting to transmit a new frame (not to retransmit a failed frame). The stationary probability distribution of the chain is $\{1/(2-p_c), (1-p_c)/(2-p_c)\}$. We use this fact to further define Equation~\ref{eq:exp:penalty} which represents the expected contention window for backoff with penalty.

\scriptsize
\begin{equation}
\label{eq:exp:penalty}
\begin{split}
E[CW] = \frac{1}{2-p_c}\sum_{i=0}^{k-1}\frac{(1-p_c)(CW_{min}-1)p_c^ir^i}{2(1-p_c^k)}  \\
+ \frac{1-p_c}{2-p_c}\sum_{i=0}^{k-1}\frac{(1-p_c)(CW_{min}-1)p_c^ir^{k-1}}{2(1-p_c^k)} \\
=\frac{1}{2-p_c}\frac{(1-p_c)}{(1-p_c^k)}\frac{(CW_{min}-1)}{2}\Bigg(\frac{(p_c^kr^k-1)}{(p_cr-1)} - r^{k-1}(p_c^k-1)\Bigg) 
\end{split}
\end{equation}
\normalsize

\begin{table}[ht]
\caption{Optimal values of $r$ for backoff with penalty and rollback backoff protocols} 
\begin{center}
\begin{tikzpicture}
\small
\node (table) [inner sep=0pt] {
\begin{tabular}{c|c|c|c|c}
N & $CW_{fixed}$\footnotemark &$E[CW]$ & {\small With penalty, $r_{opt}$} & {\small Rollback, $r_{opt}$}\\\hline
\rowcolor{Gray}2 & 12.4 & 14.9 & 1.18 & 1.11 \\\hline
               3 & 21.3 & 27.3 & 1.35 & 1.25 \\\hline
\rowcolor{Gray}4 & 30.1 & 40.1 & 1.45 & 1.31 \\\hline
	       5 & 38.9 & 55.2 & 1.53 & 1.38 \\\hline
\rowcolor{Gray}6 & 47.6 & 71.2 & 1.65 & 1.45 \\\hline
	       7 & 56.4 & 88.8 & 1.67 & 1.5 \\\hline
\rowcolor{Gray}8  & 65.1 & 107.8 & 1.73 & 1.55 \\\hline
	       9 & 73.8 & 128.5 & 1.78 & 1.65 \\\hline
\rowcolor{Gray}10 & 82.87 & 150.8 & 1.85 & 1.67 \\\hline
	       11 & 91.2 & 174.8 & 1.88 & 1.69 \\\hline
\rowcolor{Gray}12 & 100 & 200.5   & 1.95 & 1.75 \\
\end{tabular}
};
\draw [rounded corners=0.8em] (table.north west) rectangle (table.south east);
\end{tikzpicture}
\end{center}

\label{tbl:ifs}
\end{table}

{\it The optimal expected contention windows}. By defining the probability of transmission attempt $p_t$ as:

\begin{equation}
p_t=2/(E[CW]+1)
\end{equation}
similarly to~\cite{Deng:04:anew,duda:sigcomm:2005} we can further define other probabilities: $p_n$, $p_s$, $p_c$. The probability that none of $N$ competing hosts will send a packet $p_n$ is:

\begin{equation}
p_n=(1-p_t)^N
\end{equation}

The probability $p_s$ that the host will successfully transmit a packet depends on the probability that the host will attempt to send a packet and none of other $N-1$ hosts will do so. It is easy to see that such probability can be defined as follows:

\begin{equation}
p_s={N\choose 1} p_t (1-p_t)^{(N-1)}
\end{equation}  

Lastly, the probability $p_c$ that at least two hosts will send a packet simultaneously, \ie collision will occur, can be found by subtracting both $p_s$ and $p_n$ from the total probability:

\begin{equation}
\label{eq:coll:prob}
p_c=1-p_s-p_n=\sum^{N}_{i=2}{N\choose i}p_t^i(1-p_t)^{(N-i)}
\end{equation}

\footnotetext[3]{These values are identical to those that were used in prior work}

With these at hand, throughput can be defined as a function of probability $p_t$:

\begin{equation}
F(p_t) = \frac{E[Bytes]}{E[Time]} = \frac{Sp_s}{p_st_s + p_ct_c+p_nt_n}
\end{equation}

The values $t_s$, $t_c$ and $t_n$ respectively represent: The time to send a packet of length $S$ $$t_s=DIFS + TX(K) + SIFS + RX(ACK)$$ The time to detect a collision $$t_c = DIFS + TX(K) + SIFS$$ The time during which the network is idle $$t_n=Slot\ duration$$

  
The function $F(p_t)$ is strictly concave in the interval $p_t \in [0, 1]$, and thus by finding its first derivative and equating it to zero we can obtain the value of $E[CW]$ which yields the maximum throughput:

\begin{equation}
\label{eq:opt:cw}
\frac{Np_t - 1}{(1-p_t)^N} = \frac{t_n-t_c}{t_c}
\end{equation}

{\bf \it Numerical results}. To calculate the theoretical results we use $t_s,t_c$ and $t_n$ which correspond to 802.11g standard working at its maximum rate: $t_s=3.22\cdot10^{-4} s$, $t_c=2.92\cdot10^{-4} s$ and $t_n=9\cdot10^{-6} s$. To compute $t_s$ and $t_c$ we have used packet size $S=1540$ bytes. We have used Newton's method to solve the Equation~\ref{eq:opt:cw} numerically for several values of $N$. Next, we have used the obtained values of $E[CW]$ and the similar method to find the roots of Equation~\ref{eq:exp:rollback} and Equation~\ref{eq:exp:penalty}. The computed results for the first $12$ $N$ are shown in Table~\ref{tbl:ifs} and for the first $50$ $N$ in Figure~\ref{fig:theoretical1}. These results represent the optimal values of $r$ for the above values of $t_s,t_c$, $t_n$ and $S$. 

Next, we compare the empirical results obtained in experiments with the model seeded with parameters of packet size, average transmission time and time to detect a collision extracted from the empirical results. To obtain an adequate comparison of the empirical results with the model one needs to have a good estimate of the mean packet size and the time to detect a collision $t_c$. Observe that $t_c$ itself depends on both the packet size and the rate used to transmit the packet. First, we have used our data sets to compute the mean packet sizes. It appeared that for all experiments involving different $N$ and $r$ this mean was close to $1000$ bytes. Second, across all the data we have calculated the average rate, and using $1000$ bytes as the packets size we have found that the average $t_c$ was close to $366 \mu s$ for backoff with penalty and $345 \mu s$ for rollback backoff. 
And finally, using these values as parameters for the model, we have computed the numerical results and compared those against the experimental data (for optimal $r$). The results of the comparison can be seen in Figure~\ref{fig:theoretical2} and Figure~\ref{fig:theoretical3}.
The empirical curves strongly resemble the theoretical curves, especially for the case of backoff with penalty.

%% file: protocol.tex
\section{Backoff factor selection protocol}
\label{sect:protocol}

In this section we present the design, implementation and evaluation of the 
protocol which allows access points to estimate the number of active wireless 
stations to adapt the backoff factors accordingly in a dynamic way.

\begin{table*}[th]
\caption{Comparative analysis of backoff protocols} 
\begin{center}
\begin{tikzpicture}
\node (table) [inner sep=0pt] {
\begin{tabular}{c|c|c|c|c}

\backslashbox[10pt][l]{\bf Property}{\bf Protocol} & Standard & Fixed CW & Rollback & With penalty                      \\\hline
\rowcolor{Gray}Short-term fairness                                &  \it Poor      &\it Close to perfect& \it Nearly perfect   & \it Nearly perfect  \\
Throughput                                         &  \it Medium & \it Medium   & \it High      & \it High          \\ 
\rowcolor{Gray}Hidden stations (fairness)                         & \it Little   &  -   &good for $r \ge 1.6$&good for $r \ge 1.6$                      \\
Energy efficiency/Collisions                       &  \it Low/High  &  \it Low/High   &  \it High/Low  &  \it  High/Low \\
\end{tabular}
};
\draw [rounded corners=0.8em] (table.north west) rectangle (table.south east);
\end{tikzpicture}
\end{center}

\label{tbl:comparison}
\end{table*}

{\it Algorithms}. We chose two different ways to calculate the number of active wireless stations within the access point's vicinity. The first approach is a simple {\em threshold-based} adaption algorithm. In this algorithm access point counts as active only those associated stations that were transmitting at least $\tau$ units of time within the update interval $T$. $$N_{active}=\sum_{\forall i}I(\tau_i \ge \tau)$$  where $I(\cdot)$ is an indicator function mapping logical expression to 1 if it is true, and 0 otherwise 
and $\tau_i$ is the fraction of time a station $i$ was using channel during interval $T$. This algorithm is really easy to implement yet it has its drawbacks: possible errors in estimations if selected $\tau$ is too small or too large.


\begin{figure}[ht]
    \centering
    \includegraphics[width=2in]{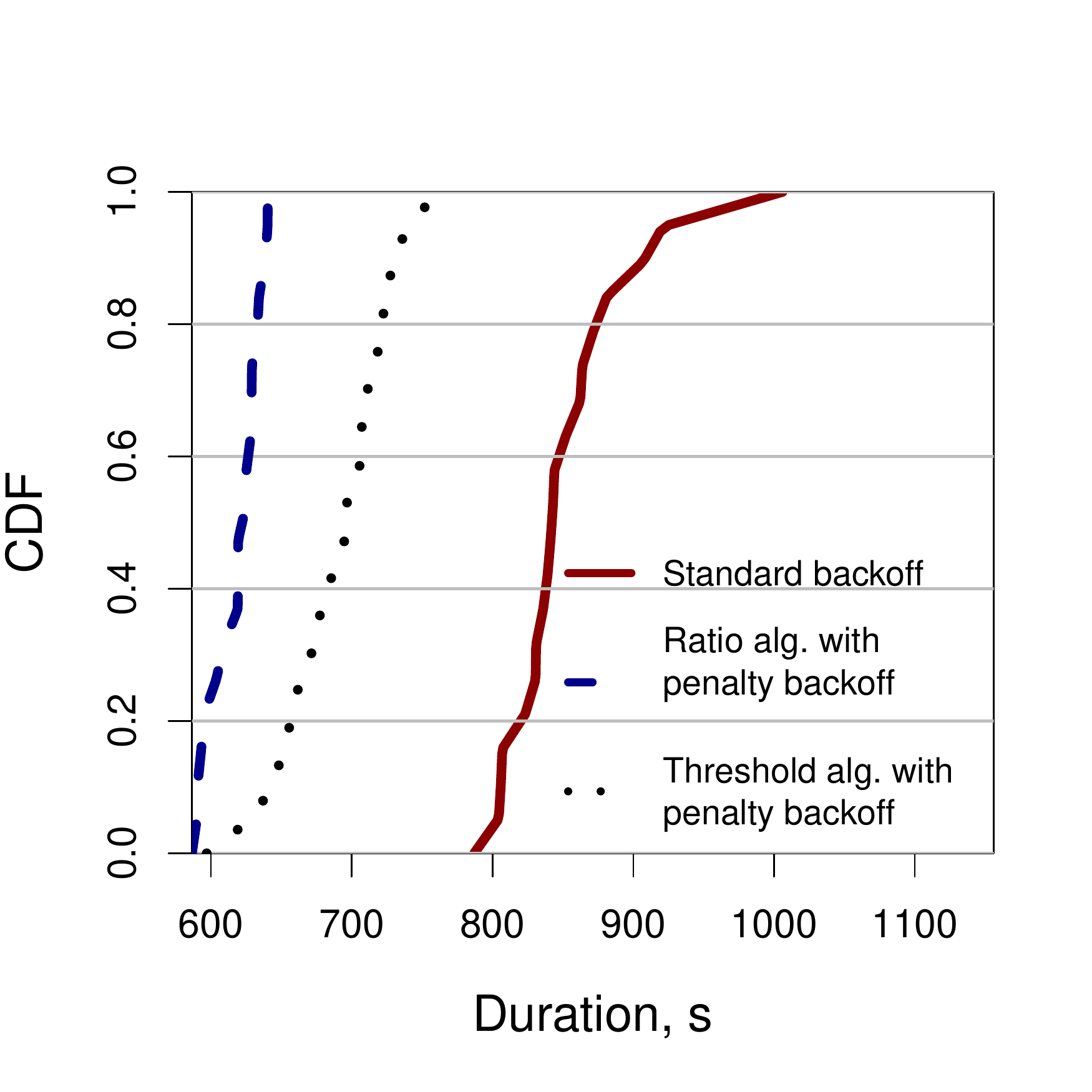}
    \caption{Comparison of experiment durations for standard backoff with penalty backoff for which backoff factors adapted using threshold- and ratio-based algorithms correspondingly}
    \label{fig:ap:selection:cdf}
\end{figure}

The second algorithm is {\em ratio-based} estimation of number of active stations. Similarly to previous algorithm, here we let $\tau_i$ to be the fraction of time during which the station $i$ was occupying the channel within the interval $T$ and define  $x=\frac{\sum_{\forall i}\tau_i}{N_{assoc}}$ to be the fair share channel access time for all associated stations. With this information at hand, we formally can define the number of active stations as follows $$N_{active}=\sum_{\forall i}I(\tau_i \ge x) + \lfloor \frac{\sum_{\forall j}\tau_jI(\tau_j < x)}{x} \rfloor$$ where $I(\cdot)$ is an indicator function. Because this algorithm can be prone to frequent oscillations in estimates of $N$ (\eg if $x$ is taken exactly equal to fair share), we introduce a coefficient $\epsilon \in (0, 1]$ and instead of $x$ we use $\epsilon \cdot x$ (we use $\epsilon = 0.8$ in our experiments). This allows the access point not to update the stations with new $N$ if the difference between between channel use is marginal. 
Finally, we note that to calculate $\tau_i$ for each station, an access point for every packet received from user $i$ stores the size of the frame  ($P_{i,k}$) and the bit rate ($R_{i,k}$) used to transmit $kth$ frame. And when time $T$ elapses for each wireless station access point recalculates $\tau_i=\sum_{\forall k}P_{i,k}/R_{i,k}$.

{\it Implementation}. We implemented both algorithms in {\it hostapd damon} - a piece of user-space software available in Linux that implements wireless access point functionality. We also introduced a new 802.11 management frame which the access point broadcasts every time it finds new value for the number of active stations. In turn, wireless clients use the value ($N$) conveyed in this frame to select proper current backoff factor. Finally, we also modified the b43 Linux kernel driver for Broadcom cards which we used in our experimental work (described in previous sections) to enable dynamical adaptation of backoff factors on wireless cards.

{\it Evaluation}. To evaluate the performance of the proposed algorithms we use the deployment in which wireless stations where scattered around the office (see Figure~\ref{fig:office:map}). The idea of the experiment is to vary the number of stations transmitting simultaneously and let the access point specify the backoff factor which the wireless stations need to use.  In this setting we use all $12$ stations. Thus, we instruct 6 stations to transmit a $40KB$ UDP stream every other $30$ seconds during $30$ seconds interval. The other 6 stations transmit $64MB$ file using TCP protocol. We repeat the experiment $20$ times for each of the following settings: (i) all stations and access point where configured with standard backoff protocol, (ii) all stations configured with penalty backoff and access point selects backoff factor using threshold-based algorithm, and (iii)  all stations configured with penalty backoff and access point selects backoff factors using ratio-based algorithm.  We show the distribution of experiment durations for all settings in Figure~\ref{fig:ap:selection:cdf}. Clearly, the {\em ratio-based algorithm} shows better performance than {\em threshold-based} algorithm and far better than standard backoff protocol which is expected. 

%% file: discussion.tex
\section{Discussion}
\label{sect:discussion}

In this section we summarize the key observations we have made during our experimental work. These observations are highlighted in Table~\ref{tbl:comparison} and explained in the proceeding paragraphs. 

{\it Does the penalty mechanism work: short-term fairness as an indicator.} It is a well-known fact that 802.11 standard backoff protocol lacks short-term fairness: nodes in a disadvantageous position fail to compete for the channel as they advance their contention window and hence decrease own odds for occupying the shared medium. Using this fact as a starting point we have introduced two modifications (which have game theoretic roots) to the standard backoff protocol: a penalty and incentive mechanisms. Two have a common goal -- improve fairness -- but tackle it in different ways. In the first one, successful nodes are penalized with larger contention windows to give a possibility for unsuccessful nodes to compete for the channel. Inversely, in the second protocol, unsuccessful nodes are rewarded with smaller contention windows. Our experiments demonstrate that these two changes can improve fairness of 802.11 networks greatly. On the other hand, unlike other studies~\cite{duda:conext:2006,duda:sigcomm:2005}, our empirical evidence indicates that simply using backoff with fixed contention windows does not guarantee good short-term fairness.

{\it Improved throughput.} In many settings we have tried, standard 802.11 protocol does achieve decent throughput. However, such performance is stipulated by capturing the channel by a small fraction of stations. In other words, to some extent, the increase in throughput (when 802.11 standard backoff is used) is a result of decrease in fairness.

On the other hand, when backoff factor $r$ is selected properly, backoff with penalty shows amazing results due to drastically decreased collisions. In particular, even in lossy environment the average improvement reached almost $100\%$ in comparison to standard 802.11 protocol. These results are closely followed by the results of rollback backoff protocol. And unlike standard backoff protocol, such improvement was not penalized with lack of fairness.

{\it Impact on energy consumption.} The literature devoted to the problems of energy efficiency in wireless networks is so extensive that we outline few out of many techniques. First, one way to reduce energy consumption by wireless devices is to accurately schedule data transmission and sleeping intervals~\cite{Xiao:2010:energy}. The other body of work discusses techniques that allow to reduce the overall amount of packets, \eg using compression and packet aggregation~\cite{Barr:2003:compression, Sharafeddine:2011:compression}. 

Although not specifically designed for being energy efficient, the two protocols proposed in this paper implicitly achieve the similar goal. We have witnessed that for instance backoff with penalty reduces the total number of collisions from 30\% to 14\%. As a result the overall number of lost packets is also reduced (bear in mind that in 802.11 there are $7$ retries before the frame is discarded). Although we do not have a quantitative estimation of the amount of saved energy, and have not explored the relationship between collisions and energy consumption, we believe that the proposed protocols have a good potential for energy saving.

{\it Can we deal with hidden terminals?} Hidden terminals is a well understood problem in wireless networks. And as we have demonstrated in Section~\ref{sec:hidden}, standard 802.11 backoff protocol lacks fairness in such settings (at least when the number of stations is $2$). Potentially, such mechanism as {\it RTS/CTS} allows to avoid the problem but at the cost of extra overhead. 

Our goal, on the other head, was to understand whether penalty and reward mechanisms are sufficient to tackle the hidden terminal problem. We have found that even though for $r \le 1.5$ backoff with penalty shows little fairness, already for $r>1.5$ the protocol passes the threshold level of $0.7$, which can be considered as good fairness, good enough to give all nodes possibility to transmit packets.

Poor fairness of the protocol with penalty when $r$ is small can be a result of almost equal values of expected backoff time and time needed to successfully send a packet. 
This can also mean that it is very likely that one of the stations will start sending a packet while the other one still hasn't finished doing so. Of the two colliding packets the access point will receive the one sent by the station with higher transmitting power, resulting in the other station consistently having a smaller fraction of successfully transmitted packets. Ideally, for the scheme to work time needed to transmit a packet should be considerably smaller than the expected backoff time. Indeed, we have witnessed that backoff with penalty achieves good fairness already when $r=1.6$ (which can be selected as minimum for real deployments as nodes will not loose much in terms of throughput).


{\it Adapting the backoff factor and estimation of active stations.} In standard 802.11 backoff protocol throughput grows with the increase of $r$ (which is however severely penalized with reduced fairness) and the trend holds for all $N$ we have experimented with. Though as our results suggest, relation between throughput and $r$ is different for backoff with penalty protocol, where an optimal value of $r$ has to be selected for each $N$. 
Indeed, setting $r$ too large when $N$ is small will result in a large number of wasted slots; choosing a too small $r$ when $N$ is large -- in large number of collisions. In Section~\ref{sec:analytical} we have demonstrated that $r$ should be a function of $N$.

In practice, however, counting the number of active stations in a distributed way can be a challenging task. For instance, in Idle Sense access protocol~\cite{duda:sigcomm:2005} the authors suggested to count the number of idle slots between any two transmissions. And it is these idle slots that are used as an indication to increase or decrease the window size. There is, however, little experimental evidence that this approach will work well in real-world deployments. For example, when hidden terminals are present, the nodes can easily miscount the number of "occupied" slots. As an alternative, the nodes can use the information provided or relayed by an access point. In this work,  we demonstrate how access points can estimate the number of active stations and broadcast this information to attached nodes. In a mesh network deployment there has to be some other distributed algorithm to find the number of active neighbors in the vicinity of a particular node. Unfortunately, we are unaware of any prior work that considered estimating number of active stations in mesh and ad-hoc wireless networks.

{\em Deployment}. We foresee at least two possible deployment paths. First, our protocol can be deployed on the nodes and coexist in a way similar to $802.11b$ and $802.11g$ protocols: If all associated stations support modified version of backoff protocol, then they all use it. Otherwise, if at least one legacy host enters the network, all nodes fallback to unmodified backoff protocol. Second, the protocols we discuss can be readily deployed in such networks where all nodes are guaranteed to follow it. For instance, in mesh networks backbone nodes can use modified backoff protocol to communicate between each other, and use unmodified backoff for communication with legacy clients (obviously using a different wireless channel).

{\it A tussle with greedy and unfair: can penalty mechanisms coexist with greedy protocols such as standard 802.11?} An interesting situation arises when, for instance, the backoff with penalty is deployed in the environment where greedy protocols prevail. Intuitively, the more greedy standard backoff can easily capture the channel, because it doesn't feature self-punishment for successful transmissions. On the other hand, the nodes that follow backoff with penalty will remain fair between each other, but will inadequately self-penalize and as a result will have smaller throughput. 

We think that the study of this tussle problem deserves a closer look and can be formulated separately. But to outline, we consider it possible to find a protocol from the family of penalty-based backoffs which potentially can tackle the problem. We leave this for future work.

%% file: related.tex
\section{Related work}
\label{sec:related}
The literature on wireless communication is so extensive that in the proceeding paragraphs we can only summarize a few of the many
lines of research related to what we discuss in this paper.

{\em Modeling and simulation of IEEE 802.11 protocol}. This body of work is arguably the largest and mainly focuses on analyzing and modeling the performance of standard 802.11 protocol. Thorough review of the key results from just most notable papers would require a book on its own, therefore, we merely mention such work in this area. For example, the following papers:~\cite{bianchi:performance,duda:80211,tay:capacity:model, vu:collision:model,kumar:new, song:enhancement,song:analysis, cho:fundamentals, lukyanenko:performance, FelembanE11} can be considered as a good starting point for deeper understanding of performance issues of 802.11 wireless networks. Whereas, we use theoretical achievements found in this literature as the basis for our own analytical models. However, our study different from the prior works in at least two ways. First, we do not analyze standard IEEE 802.11 backoff protocol as such, and instead we use mathematical tools found in the papers to investigate our own designs. Second, we do not rely on simulations, but observe the behavior of protocols in deployments with real devices and environments.

{\em Optimizations not related to backoff protocol}. The literature in this area is rife with various approaches and here we outline just two research directions. For example, recently researchers focused on efficient network coding schemes to reduce collisions and increase overall system throughput~\cite{Gudipati:2011, Gollakota:zigzag, Katti:2008:coding,Katti:2007:anc,Katti:2006:xor}. Other researchers explore the possibility to utilize radio spectrum more efficiently. Examples are designs that use multiple input-output antennas~\cite{Lin:2011:RAH, Rahul:2012:JSW}. Although these mechanisms are orthogonal to our approach, they illustrate one way of coping with collisions in wireless networks. We consider these approaches to be supplementary to our work, \ie can coexist.

{\em Modifications to backoff protocol}. There exists a considerable number of work that attempt to improve throughput and fairness of wireless networks with non-standard backoff schemes. For example, the work described in~\cite{Tan:2010:fca,Starzetz:2009, duda:sigcomm:2005, kwak:performance} uses non-standard contention windows. Unlike these studies, which suggest to either remove the exponential backoff completely and use fixed contention windows or non-standard backoff factors, we take a radical step and propose to use non-standard state transitions to penalize certain stations. The work in~\cite{Celik:2010:MNM} is the closest to ours. Similarly to us, the authors suggest to use non-standard state transitions and penalize too successful stations to improve fairness and throughput. However, there are several key differences the approaches. First, unlike our work, in~\cite{Celik:2010:MNM} the authors use only two states and hence two contention windows, meaning that the backoff protocols are rather different. Second, in~\cite{Celik:2010:MNM} the authors apply their design to multipacket reception wireless networks; whereas in our designs, although we experiment with the clients using a single channel for transmitting and receiving packets, we do not limit ourselves to any underlying physical layer technology. Finally, although the conclusions we derive are similar, in our work we make our judgment  based on empirical study and do not rely on simulations.

{\em Estimating number of active stations}. There is a number of proposals on estimating number of active stations in 802.11 wireless networks. Bianchi \etal~\cite{bianchi:96:mac:perf} were the first to propose using busy slots to achieve this goal. Cali \etal~\cite{Cali98ieee802.11} investigate this direction further, and derive a metric which estimates the number of active stations based on the observed number of idle slots. Even though these metrics can be readily used in our design, their accuracy in case of uneven demands for channel resources is an open question. Neither it is clear how well these metrics will perform in real-world deployments with hidden terminals. The other body of work considers using a centralized controller~\cite{Wu:2007:PAT} in enterprise wireless networks which collects the information about used channel time and available bandwidth. This information can be used to tune wireless network parameters. These approaches do not bear into our situation directly, but can be used, for example, in large scale deployments of our designs. Finally, we are aware of only one prior work~\cite{duda:conext:2006} where the authors investigate the techniques to estimate number of active stations empirically. Thus, our work offers a considerable step towards better understanding

{\em Experimental work}. We find it surprising that there exists little empirical investigation of modified backoff protocols. As such, we have found that only in~\cite{duda:conext:2006} the authors reported some practical implementation and evaluation of backoff protocol using proprietary hardware and firmware and a small number of wireless stations. And few research works~\cite{TinnirelloBGGGG12} consider implementation of MAC protocols in general on commodity hardware. Thus, our work is another step towards better understanding of real-life performance of modified MAC protocols on commodity hardware.

%% file: conclusion.tex
\section{Conclusion}
\label{sec:conclusion}

The main goal of this paper is to evaluate performance of various 802.11 backoff protocols in real-life experiments. 
The four protocols we considered were standard backoff, backoff with fixed contention window, backoff with penalty and rollback backoff. 
The first step in our work was to implement the protocols on suitable devices and design the experimental setup.
We described in details the difficulties we had encountered during implementation of backoff protocols on commodity 
wireless cards, since we believe that these details may help other researchers who will be experimenting with
the same or similar devices.

We put considerable effort in designing our experiments and making sure that our data is consistent. Our study includes
several scenarios: in lossy and normal environment, with upload and download traffic pattern, with hidden stations and
delay sensitive applications, with TCP and UDP flows. In the experiments we observed the consequences of varying 
backoff factor from $1.2$ to $2.6$ 
and number of stations from $3$ to $12$. The collected data was carefully scrutinized and calibrated to avoid any errors. 

The three main performance metrics we have monitored were aggregated throughput, short-term fairness and collision probability. 
In our experiments standard 802.11 backoff protocol and backoff with fixed contention window demonstrated poor
fairness. We have concluded that unfair behavior is built-in to the standard 802.11 backoff protocol,
where unsuccessful stations are forced to remain unsuccessful, while successful stations are able capture the 
shared medium for long periods and remain successful.

To mitigate the problem with fairness, we have introduced self-penalty mechanism in backoff with penalty and rollback backoff
protocols. Our empirical evidence, simulation results and an analytic model reveal that these two protocols 
achieve nearly perfect fairness and throughput 
up to $100\%$ higher than in standard backoff. In addition, the penalty mechanism allowed to decrease collision rate by
more than half and to reduce delays in delay sensitive traffic.